\pdfoutput=1
\documentclass[letterpaper,twocolumn,10pt]{article}
\usepackage{usenix-2020-09}

\usepackage{url}
\usepackage{microtype}
\usepackage{tabularx}
\usepackage{soul}
\usepackage{subcaption}
\usepackage{amsthm}
\usepackage{amssymb}
\usepackage{listings}
\usepackage{tikz}
\usepackage{booktabs}
\usepackage{xspace}
\usepackage{balance}
\usepackage{multirow}
\usetikzlibrary{arrows}
\usepackage[font=small,format=plain,labelfont=bf,textfont=it,aboveskip=0pt]{caption}
\usepackage[compact]{titlesec}  

\usepackage{float}
\floatstyle{plaintop}
\restylefloat{table}

\newcommand{\TITLE}{Understanding How and Why University Students Use Virtual Private Networks }

\let\origquote\quote
\let\endorigquote\endquote
\renewenvironment{quote}{%
    \small\origquote
}%
{\endorigquote}

\begin{document}

\title{\TITLE}

\author{
    {\rm\small Agnieszka Dutkowska-Zuk, Austin Hounsel\footnotemark[1], Andre Xiong\footnotemark[1],  Molly Roberts\footnotemark[2], Brandon Stewart\footnotemark[1], Marshini Chetty\footnotemark[3], Nick Feamster\footnotemark[3]} \\
    {\small Lancaster University, Princeton University\footnotemark[1], University of California at San Diego\footnotemark[2],
    University of Chicago\footnotemark[3]}
}

\maketitle

\begin{abstract} 

    We study how and why 
    university students chose and use
    VPNs, and whether they are aware of the security and privacy risks that 
    VPNs pose. To answer these
    questions, we conducted 32 in-person interviews and a survey with 349
    respondents, all university students in the United States. We find
    students are mostly concerned with access to content and 
    privacy concerns were often secondary. They made tradeoffs to
    achieve a particular goal, such as using a free commercial VPN that
    may collect their online activities to access an online
    service in a geographic area. Many users 
    expected that their VPNs were collecting data about them, although they 
    did not understand how VPNs work. We conclude with a discussion 
    of ways to help users make choices about VPNs.

\end{abstract}



\begin{sloppypar}
    \section{Introduction}

Virtual Private Networks (VPNs)~\cite{donenfeld2017wireguard} encrypt all
network traffic from a client device to an intermediate server. As a result, 
many users rely on VPNs to access blocked content or to preserve their privacy.
Many VPN services are now available, with companies from Cloudflare to
Facebook now also provide their own VPNs~\cite{pcmag_vpn_reviews,
facebook_onavo, cloudflare_warp}. Some estimates indicate that the VPN market
has grown from \$16.5 billion in 2016 to \$20.6 billion in
2018~\cite{geosurf_vpn_usage}.  Yet, despite their relatively widespread use,
and in spite of their name, many VPNs fail to provide even basic security.
For example, some VPNs have accidentally leaked user traffic, breaking
security and privacy claims made by the providers~\cite{perta_15, khan_1}.
Other VPNs may capture user traffic and send the data to third parties for
targeted advertising~\cite{ikram_6, hotspotshield_complaint}. At one point,
Facebook Onavo collected application traffic without notifying
users~\cite{businessinsider_onavo, techcrunch_onavo}.  

In light of the fact that many VPN services are not in fact private, we aim to
better understand how and why VPN users use them, and 
whether they are aware of the security and privacy risks that VPNs pose. We also 
aim to understand how these users choose which VPNs to use. 
To answer these questions, we recruited a large, diverse population of students 
from a university in the United States. We focused on university students 
because universities typically offer a VPN to their students and employees to access
various content and services (e.g., at some universities, various compute
clusters and services are only available via VPN, even on campus).  
Thus, this population typically has access to and awareness about at
least one VPN, affording at least a basic level of familiarity with VPNs, if
not more experience with VPN usage.
We focused on the following questions: 
\begin{itemize} 
\itemsep=-2pt
\item Why do students use VPNs?
\item How do students choose which VPNs to use?  
\item Are students aware of security \& privacy risks of VPNs?
\end{itemize} 
\noindent To answer these questions, we performed a mixed methods
study with two parts (qualitative and quantitative): (1)~in-depth interviews with 32 university students, and
(2)~a large-scale survey of 349 university students. 

We found that different sets of students have differing levels of awareness and attitudes toward VPNs. Most university students used VPNs to gain access to content and
materials at their institutions (e.g., restricted pages, library materials),
or to bypass censorship or filtering of content.
Interestingly, privacy and security were secondary considerations.
Students made tradeoffs with their privacy and security to achieve a pragmatic 
goal, such as using a free VPN that may collect information about their browsing 
history to access geographically restricted content.
Indeed, many students did not expect VPNs to provide
privacy and some even expect that the provider may be collecting data about them
and even providing third parties access to that data.

We also found that some students were generally not familiar with the technical
details of how a VPN works, which often led to misconceptions and
misunderstandings about the privacy guarantees that a VPN could provide.  Some
of these misunderstandings were even more fundamental, suggesting that not
only did students not understand technically how VPNs work, but they also did
not understand the capabilities and incentives for various VPN providers to
collect data about them. For example, although many students indicated that
they used a VPN to protect their data from ``companies'' in general, they
seemed unconcerned that the VPN provider itself is a company (and, in the case
of some, such as Facebook's Onavo VPN, even an advertiser) that is often
gathering user data for profit.

Our results suggest possible future directions for helping Internet users
safely use VPNs, particularly along the lines of improving awareness about how 
data collection works and when it is happening. 
Users not only misunderstand the technical capabilities of VPNs,
but they also have sufficient information to fully appreciate how VPNs manage their privacy.
Better technical design can lead not only to more informed users, but also
instill positive outcomes where users can use that information to make better
choices about their selection and use of VPNs.

    \section{Background and Related Work}

We provide background on VPNs and survey related work, including 
work studying privacy and security vulnerabilities from VPNs and past studies
of user attitudes about privacy.

\paragraph{Background: Virtual Private Networks.}
Originally created for enterprises to communicate securely, VPNs rapidly
gained broad commercial appeal as personal Internet usage
soared~\cite{hirst_26}. VPNs create secure connection (a ``tunnel'') to a
server from which user can safely access a
destination~\cite{vaughan-nichols_25}. VPN providers can encrypt and
authenticate this connection using a number of methods with varying degrees of
effectiveness, including OpenVPN, Layer 2 Tunneling Protocol, Internet
Protocol Security, and several others \cite{donenfeld2017wireguard}. From the
perspective of a network eavesdropper, the VPN user's traffic appears to be
coming from the VPN server, as opposed to from the user's device.  VPNs can be
used to access destinations on the Internet or on private networks, such as on
a private enterprise or campus network.  Users can also use VPNs to access
blocked content, such as Twitter in China~\cite{anderson_2012} or to access
location-restricted content~\cite{nordvpn_features}.  Commercial VPN providers
often offer multiple servers~\cite{nordvpn_servers}.


\paragraph{VPN Data Leakage}
Researchers studied 14 of most popular VPN providers and found that most of
these providers unintentionally leak traffic to websites hosted on IPv6
addresses~\cite{perta_15}.  Researchers have also found that off-the-shelf VPN
software is susceptible to passive and active attacks, limiting their ability
to provide anonymity~\cite{appelbaum2012vpwns, al2017one}.  Other researchers
studied commercial VPN providers and found five providers that implement
transparent proxies, which inspect and modify traffic that users
send~\cite{khan_1}.  Researchers studied VPN apps in the Android marketplace
and discovered that many VPNs send data to third-party trackers or have
security misconfigurations~\cite{ikram_6, zhang_17}.  Researchers surveyed
Pakistani Internet users and found that 57\% of respondents used VPNs to
access YouTube while the website was censored in 2012~\cite{khattak2014look}.
Past work explored how smartphone users used VPNs
when Facebook was banned in Sri Lanka~\cite{jayatilleke2018smartphone}; our
study concerns a different demographic and is more extensive.

\paragraph{User Attitudes on Privacy}
Researchers have analyzed users' mental models in their perceptions of the
Internet~\cite{kang_2,klasnja_16, poole_17, krombholz2019if}.  Users in the United States are
concerned about online tracking and want to control it in certain
situations~\cite{melicher2016not, smith_18, turow_19,rader_8},  yet they are
often confused as to how tracking works and how they can protect
themselves~\cite{shirazi_9}.  One study suggested that a combination of
awareness of, motivation to use, and knowledge of how to use privacy and
security tools impacted their usage~\cite{das_20}.  However, another study
focused on online privacy and security attitudes and behaviors found that
while Internet users with stronger technical backgrounds were more aware of
privacy and security threats, they did not engage in more secure practices
than their less knowledgeable peers~\cite{kang_2}, with the exception of
``expert users''~\cite{busse2019}. Previous research showed that 18-to-22 year olds are
likely to rely on strategies to make themselves less visible 
online~\cite{pew-research-2016}. These phenomenon of 
tech-savvy and younger users neglecting to use their knowledge to protect themselves could
have implications for VPN-focused studies.

    \section{Method} 

We study how and why students use VPNs, their mental models of
VPNs, how they choose which VPN to use, and their
awareness and attitudes about data collection practices of VPNs.
To do so, we conducted 32 qualitative interviews \cite{Seidman}
and a large scale survey with 349 university students from one of our
institutions. Both parts of the study were approved by our institution's Institutional
Review Board~(IRB).

\subsection{Interviews} 

Before participating in a semi-structured interview, participants were asked
to fill out a consent form and a short questionnaire.
We collected their academic
majors and other basic demographic information such as age, gender, course
of study, and information about their online habits. The interview guide was structured to first get a better understanding of
participants' knowledge and background and participants' general privacy and
security awareness. 
First, we asked participants who they believed could collect data about them
online and who they would want to prevent seeing certain information about
their online habits. We then asked
participants to describe how a VPN works. Next, we asked how they learned about VPNs and what
their first experience using a VPN was. We then asked how participants choose
to use a particular VPN and how and why they use a VPN, 
how participants felt when using a VPN, and whether they use a paid or free
versions of VPNs. Finally, we asked about VPN issues and improvements,
students' knowledge and usage
patterns of different VPN types, including specific VPNs they had used,
about reasons for selecting and using VPNs,
perceptions of data collection by VPNs, and any other issues that they faced.

\subsubsection{Recruitment}
\begin{table*}[t]
\centering
\begin{tabular}{l r r |l r r|l r r|l r r} 
 \hline
 Age & \# & \% & Gender & \# & \% & Origin & \# & \% & Educational status & \# & \% \\
\hline
18 to 24 & 26 & 81\% & Female & 17 & 53\% & United States & 12 & 37.5\% & Postdoctoral Researchers & 4 & 13\% \\
25 to 34 & 5 & 16\% & Male & 14 & 44\% & International & 20 & 62.5\% & Graduate students & 2 & 6\% \\
35 to 44 & 1 & 3\% & Other & 1 & 3\% & & & & Undergraduate students & 26 & 81 \% \\
 \hline
\end{tabular}
\caption{The distribution over age, gender, origin and education status for 32
    interview participants, at the time of collecting the data; 20 international participants came from 17 different countries.}
\label{table:1}
\end{table*}

We recruited 32 interview participants via an institutional survey research center and social media accounts such as Twitter. Table~\ref{table:1} shows demographic
data of the interview participants, who were mostly 18--24 years old and undergraduate students (81\%). 
We filtered for students who had used a VPN before, and for students that are currently enrolled in a particular United States university's undergraduate or graduate
program. We aimed to recruit a variety of international and domestic students
living in the U.S. We concluded that such diverse group would
expand our knowledge and understanding on how and why participants use VPNs.
Interviews were conducted in Summer and Fall 2018. Participants were
compensated with a \$20 Amazon gift card.  
We conducted 23 interviews via Skype, and another nine were conducted on a university
campus. Four interview participants did not give consent to recording, so detailed notes
were taken during these interviews. All other interviews were audio recorded.

\subsubsection{Data Analysis} 

We first transcribed all recorded interviews and developed an extensive codebook
to apply to the interview transcripts and field notes. We used the Dedoose
platform~\cite{dedoose}
for qualitatively coding the transcripts for a thematic analysis \cite{saldana_coding_2013}.  
One of the research team first coded all of the interview transcripts and a second member of the team reviewed the coded transcripts for consistency. When there were any points of disagreement, this was discussed and resolved as a team. We had 45 parent codes, each with several child codes, for a total of 1906 codes. Once the transcripts were all coded, the whole research team met and discussed the codes to identify 9 parent codes of interest, shown in Table \ref{table:3}, with 31 child codes as reflecting the main themes in the data. For example, our main code \textit{Reasons for VPN usage} had four child-codes, which represented participants' motivation behind using VPNs: \textit{Bypass geographic firewalls, Work, Privacy, Not Privacy/Security}. To illustrate, an example quote linked to \textit{Bypass geographic firewalls} child-code was participant's 26 testimony: \textit{`I have my mom reroute the US IP address
to a Mexican IP address with a VPN, so then she could watch her Venezuelan TV shows.'}

The primary coder then wrote summaries of these codes. The research team
reviewed the summaries and held regular research meetings to decide on the
final themes arising from the interview data. We also used email and a
dedicated Slack channel to communicate about the paper themes. 
Calculating inter-rater reliability (IRR) is not necessary for the type of
analysis that we performed, because shared consensus can still be reached
without this measure in thematic analysis~\cite{armstrong1997place}.  McDonald
et al. also state that calculating IRR is not necessary when coded data is
not the end product but instead part of the process to derive concepts and themes, or in our case, as input for thematic
analysis~\cite{McDonald_2019}.  

  \begin{table*}[t!]
  \small
    \centering
    \begin{tabular}{|l |l|} 
    \hline \textbf{Main Code} & \textbf{Definitions} \\
     \hline
    {Reasons for VPN usage} &    
   motivations and goals behind VPN usage
    \\
    {What is a VPN?} &    
    mental model of a VPN
    \\
   {Guidelines when choosing VPN} &     personal preferences of VPN's qualities and strategy in choosing a VPN \\
   {Trust in VPN provider} &   
     attitudes towards their VPN provider \\
    {Using institution related VPN} &  
         usage habits regarding institutional VPNs  \\
   {Use or trust free VPN} &   
      attitudes towards commercial free VPNs \\
   {VPN practices} &   
   thoughts on VPNs' data practices \\
   {What a VPN guarantees} &    
    assumptions about VPNs' assertions \\
{Tracking while using VPN} &   
    assumptions about protection that VPNs provide against tracking  \\
     \hline
    \end{tabular}
    \caption{Summary of main codes reported in the interviews.}
    \label{table:3}
    \end{table*}
    
\subsection{Survey} 

Based on the interview data and analysis, we then designed a larger-scale survey to complement our interviews data and expand our knowledge about VPN users' perspectives. We first pre-screened and
filtered out respondents who did not consent to the survey, were under 18, or
had never used a VPN. As in the interviews, we collected academic
majors and other basic demographic information such as age, gender, and course
of study. We also collected background information about respondents' perceptions
and concerns about data collection, including the nature of the data
collected, who is collecting data, and why they are collecting data. We also asked about respondents' usage patterns
of different tools and tactics to combat online risks, as well as how they
sourced them.

We asked similar questions as in the interviews, but we generally avoided
open-ended questions to prevent user fatigue and reduce the complexity of data
analysis; as a result, we asked only three open-ended questions. We also avoided
double-barreled questions, negative questions, and biased wording
\cite{lazar_28}. We included two attention check questions that required a
certain response to ensure respondents were answering mindfully. Participants that had been interviewed in the first part of our study were not allowed to take the survey, to avoid response bias.

\subsubsection{Recruitment}
\begin{table*}[t]
\centering
\begin{tabular}{l r r |l r r|l r r|l r r} 
 \hline
 Age & \# & \% & Gender & \# & \% & Origin & \# & \% & Educational Status & \# & \% \\
\hline
18 to 25 & 274 & 79\% & Female & 178 & 51\% & United States & 257 & 74\% & Graduate Students & 123 & 35\% \\
26 to 35 & 74 & 21\% & Male & 171 & 49\% & International & 92 & 26\% & Undergraduate Students & 226 & 65\% \\
36+ & 1 & 0\% & & & & & & & & & \\
 \hline
\end{tabular}
\caption{The distribution over age, gender, origin, and education status for 349 survey participants, at the time of collecting the data. Our 92 international participants came from 32 different countries.}
\label{table:2}
\end{table*}

We recruited undergraduate and graduate students from a 
large university in the United States to take the survey on Qualtrics. We sent email invitations to a
random sample (containing 2,748 people) of the university population via an institutional survey
research center. We aimed to reach at least 5\% of the university's VPN-using student population as recommended by Lazar \cite{lazar_28}. 
We launched and conducted
the survey between February 2019 and March 2019. Our large sample size allowed us to collect 452 responses, of which 349 were completed, passed
our attention checks, and fit our recruiting criteria. Our final sample of 349 valid and completed responses is
large compared to university's overall population (4.3\%). Table~\ref{table:2} shows detailed demographic data of the respondents. As with the interviews, the
majority of them were age 25 and under (79\%). Participants with complete valid responses were entered into a draw for one of two \$250 Amazon gift cards. 

\subsubsection{Data Analysis} 

We used Qualtrics and R to analyze the survey data. We first analyzed the
response data using tools built-in with Qualtrics. We limited our analysis to
the 349 valid and complete responses. First, we performed descriptive analysis on all the survey questions. The respondents were required to answer every question except for Figures \ref{fig:q5_4_Paid_commercial} and \ref{fig:q5_5_Free_commercial}, but certain questions were only shown when applicable. As such, questions that have
fewer than 349 data points contain responses from every applicable respondent;
a lack of response does not indicate a respondent's choice to abstain, unless it was an open-ended question. In
presenting our results, we show counts in terms of how many participants were shown a question.

We also include counts for those who were not shown the question, or chose not to answer in the case of Figures \ref{fig:q5_4_Paid_commercial} and \ref{fig:q5_5_Free_commercial}.
We qualitatively coded the open-ended answers in a similar fashion
to the interviews using a code book that was developed based on multiple reads
through the responses. The graphs show all survey codes. One team member coded all the responses but a second team member read the coded responses to ensure consistency and to discuss any points of disagreement. Since the coding pass was primarily done by the first researcher, we did not calculate IRR. In the graphs, we report on the most prevalent codes occurring in the survey data. In the graphs presented, response count reflects the
total number of participants who chose an option, oftentimes this was in
answer to a ``Check all that apply'' question, so the total of all the
responses may be greater than 349 if any participant selected multiple
options. In questions where participants were asked to choose and rank
options, we compute a weighted score on the inverse ranking, where weights
correspond to $1/r$ for a ranking of $r$~\cite{stillwell1981}. 

We also searched pairwise correlations across the results of each survey
question. We reduced the number of observed variables by consolidating
questions into fewer categories when applicable. Given the broad nature of
this study, we were unable to account for unpredictable human
behavior and the many unobserved variables that substantially impacted the
practicality of this analysis. As such, we expected the pairwise correlations
to produce numerous significant values (p < 0.05), but with low R. The tests
confirmed our hypothesis, but the correlations were not sufficiently
meaningful to report.

For certain survey questions, we show additional figures for meaningful
differences in responses between all participants (N=349) and participants
that only use VPNs provided by their university (N=49).  For these same
questions, we summarize the responses for participants that only use
commercial VPNs (N=91), and we provide Spearman correlation coefficients
comparing all participants to participants that only use commercial
VPNs~\cite{spearman1987proof}.  Spearman correlation coefficients enable us to
measure the association between the ranks of responses given by two groups of
participants (1 indicates a complete positive association (identical ranks);
-1 indicates complete negative association.  For other survey questions, we
did not find meaningful differences in responses between the two groups, so we
do not show additional figures. When reporting our qualitative data, we refer
to survey participants' as ``S'' and to interview participants as ``P''.

\subsection{Limitations}

Our study has several limitations. First, we focused on students at one
university in the United States.  Although the university makes a concerted
attempt to recruit a diverse cross-section of students, any single university
may not be representative of all university students.  Second, we only focus
on university students, rather than other populations of VPN users. Other
populations may have different attitudes and behaviors, but we do not
generalize our results to them.  Future research could replicate this study
with other populations (will release our survey and interview scripts).
Second, our survey and interviews also have limitations.  Recall bias is
difficult to avoid in any survey~\cite{lazar_28}. Our survey and interviews
were not completely anonymous as they required survey participants who wished
to enter the raffle and all interview participants to submit an email.
Interview participants were asked to meet with one of the research team in
person which could affect the respondents' honesty.  Third, our survey did not
distinguish between precise notions of safety and privacy, thus
participants could interpret them differently. 

\begin{figure}[t]
    \centering
    \begin{subfigure}[t]{0.49\textwidth}
        \includegraphics[width=0.9\columnwidth]{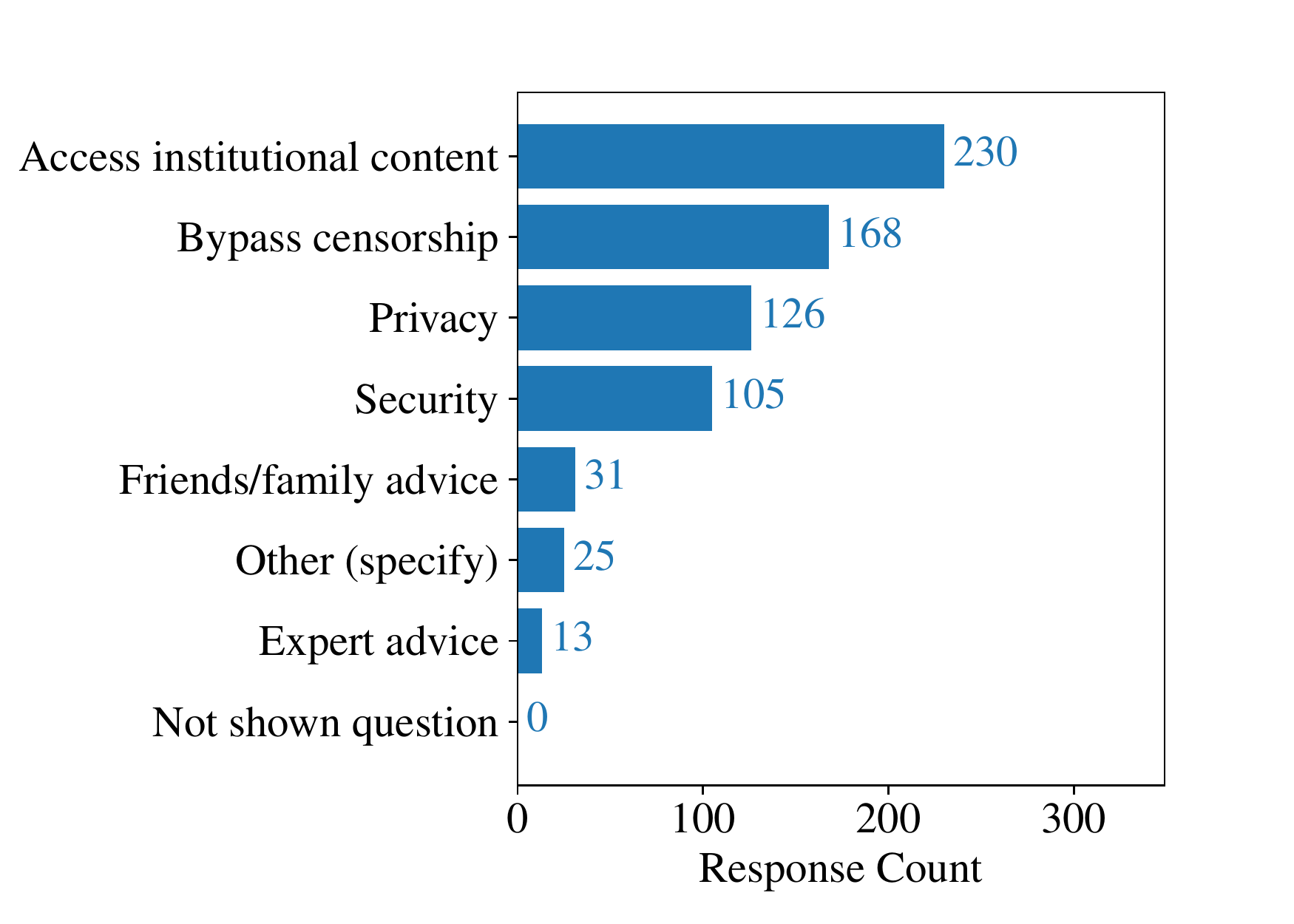}
        \caption{All participants.}
        \label{fig:q5_17_all_participants}
    \end{subfigure}
    \hfill
    \begin{subfigure}[t]{0.49\textwidth}
        \includegraphics[width=0.9\columnwidth]{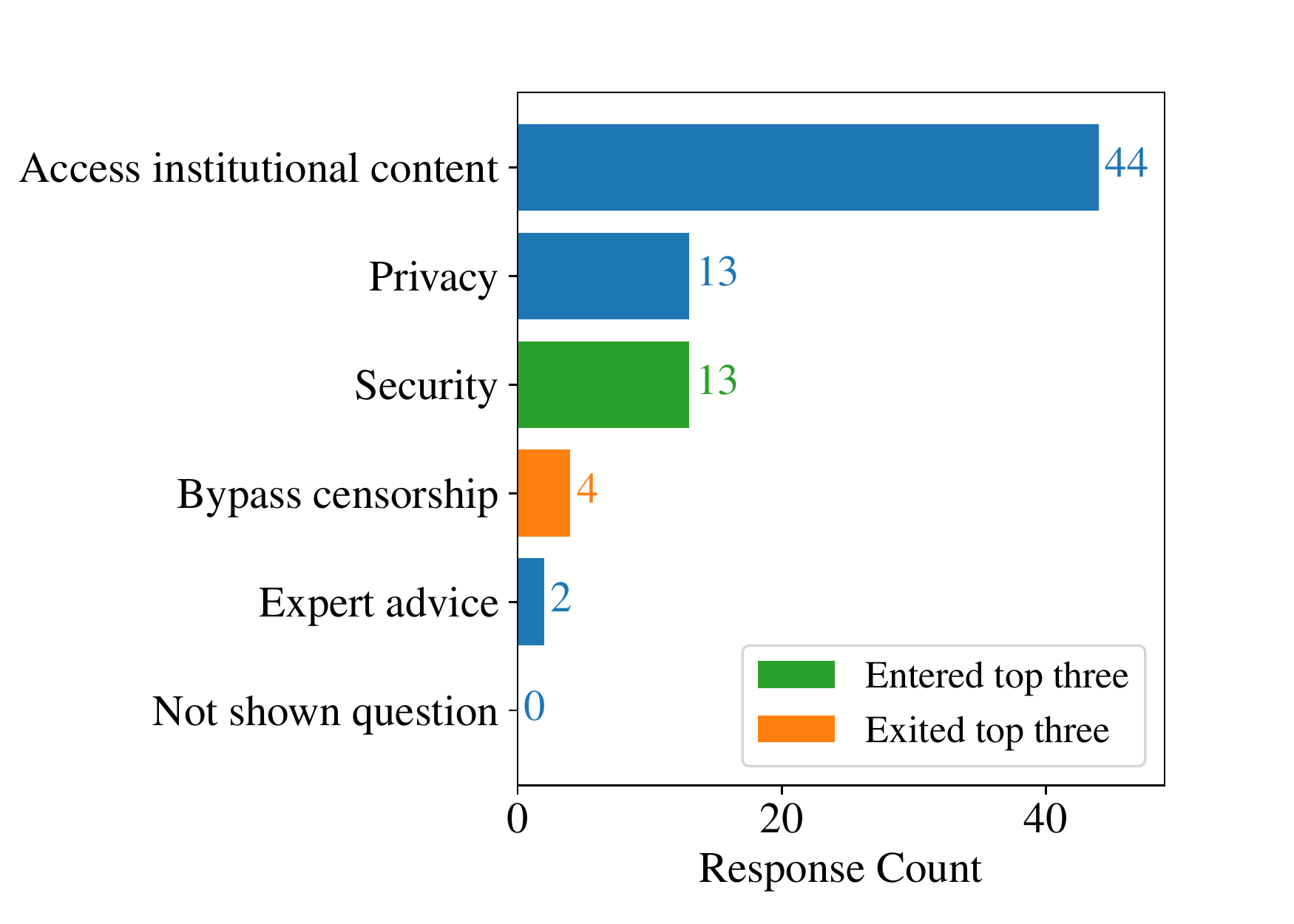}
        \caption{Participants that only used university VPNs.}
        \label{fig:q5_17_university_vpn_only}
    \end{subfigure}
    \caption{Why do/did you use a VPN? (responses selected by participants)}
\end{figure}

    \section{Results}\label{sec:findings} 

We present findings on on how and why students used VPNs, their mental 
models of VPNs, how they chose VPNs, and their 
attitudes about data collection by VPNs.

\subsection{How and Why Students Use VPNs}

We found that although privacy is of some concern for students, it is often
secondary consideration. Furthermore, most students did not use VPNs regularly. 
When they did use VPNs, it was mainly to access blocked content and
institutional materials, not to protect privacy. 

\subsubsection{Students mostly use VPNs to access content}

\paragraph{Interview}
Most interview participants (21/32) reported that they used a VPN to bypass geographic firewalls, and to
watch movies or TV shows online (15/32). For eleven interview participants,
accessing blocked content was the priority when using a VPN.
In a typical example, P11 spoke of using a VPN for news websites that were not blocked but had different or
limited content depending on IP address of the Internet user. As they were
from United Kingdom (UK), they wanted to access the UK BBC website while they were in the
US. Also, P26 shared that they used a VPN to help their mom: 
\begin{quote}Venezuela has blocked everything coming from their YouTube channels, and I have my mom reroute the US IP address to a Mexican IP address with a VPN, so then she could watch her Venezuelan TV shows.\end{quote}
Nevertheless, interviewees would use VPNs to resolve different issues, as P24 explained: \begin{quote}I found out that I couldn't access the application or login to my account
through the phone application, because I wasn't in the United States(\dots) And I downloaded, I
think it's called Express VPN. And that was able to help me work around the
location, geographical issue, and access the account so I could cancel the
subscription.\end{quote}

\paragraph{Survey}
Most survey respondents also used a VPN to access content, specifically institutional materials
when off campus~(230/349)~(Figure~\ref{fig:q5_17_all_participants}).
Additionally, 168/349 survey respondents reported using a VPN to bypass Internet
censorship. However, 138/349 survey participants disclosed that they used VPNs to protect privacy or security. 

For those who specified ``Other'' in the survey, students
commonly reported using a VPN to access Advanced Placement (AP) scores, as
S108 noted: ``To
access AP scores early (they were releasing them one time zone at a time to
prevent too much web traffic)''. 

Most students who only used VPNs provided by their university used them 
to access institutional materials (44/49). A few students used university-provided 
VPNs to access blocked content (4/49)
(Figure~\ref{fig:q5_17_university_vpn_only}). 
Conversely, most students that only used commercial VPNs used them 
to access blocked content (55/91). Interestingly, some of these students used 
commercial VPNs to access institutional materials (23/91).
For this survey 
question, there was also a significant correlation between the responses given 
by all participants and participants that only used commercial VPNs, with a 
Spearman correlation coefficient of 0.786 (p = 0.036).

\subsubsection{Fewer students used VPNs for privacy \& security}
\paragraph{Interview}
Thirteen interview participants said that privacy was not the main reason for
using a VPN.
Fewer (7/32) used it to protect their personal information, and four wanted a VPN to be secure and keep them anonymous, such as, P21: \begin{quote}I guess I don't like
    the idea of [the university] or an ISP being able to see all of my traffic. I don't
think that I trust anyone with all of my traffic or consumer habits.\end{quote}
A few interview participants (3/32) used VPNs because they liked the idea that there was a
{``free''} space on the Internet. For these types of participants, using a VPN could be a strong statement
that privacy is important, as P25 explained:

\begin{quote}It's why Private Network Access got so
popular. They tried to subpoena the guys to release information about some of
the people who used the VPN, and then they actually didn't have it on their
servers. So people knew that they didn't keep records, so everybody started
using that one. \end{quote}
Yet, some interview participants trusted VPNs more than other networks. For instance, five participants said they would use a VPN while on public Wi-Fi and
four while traveling.

\begin{figure*}[t]
    \begin{minipage}[t]{0.33\linewidth}
        \centering
        \includegraphics[width=\linewidth]{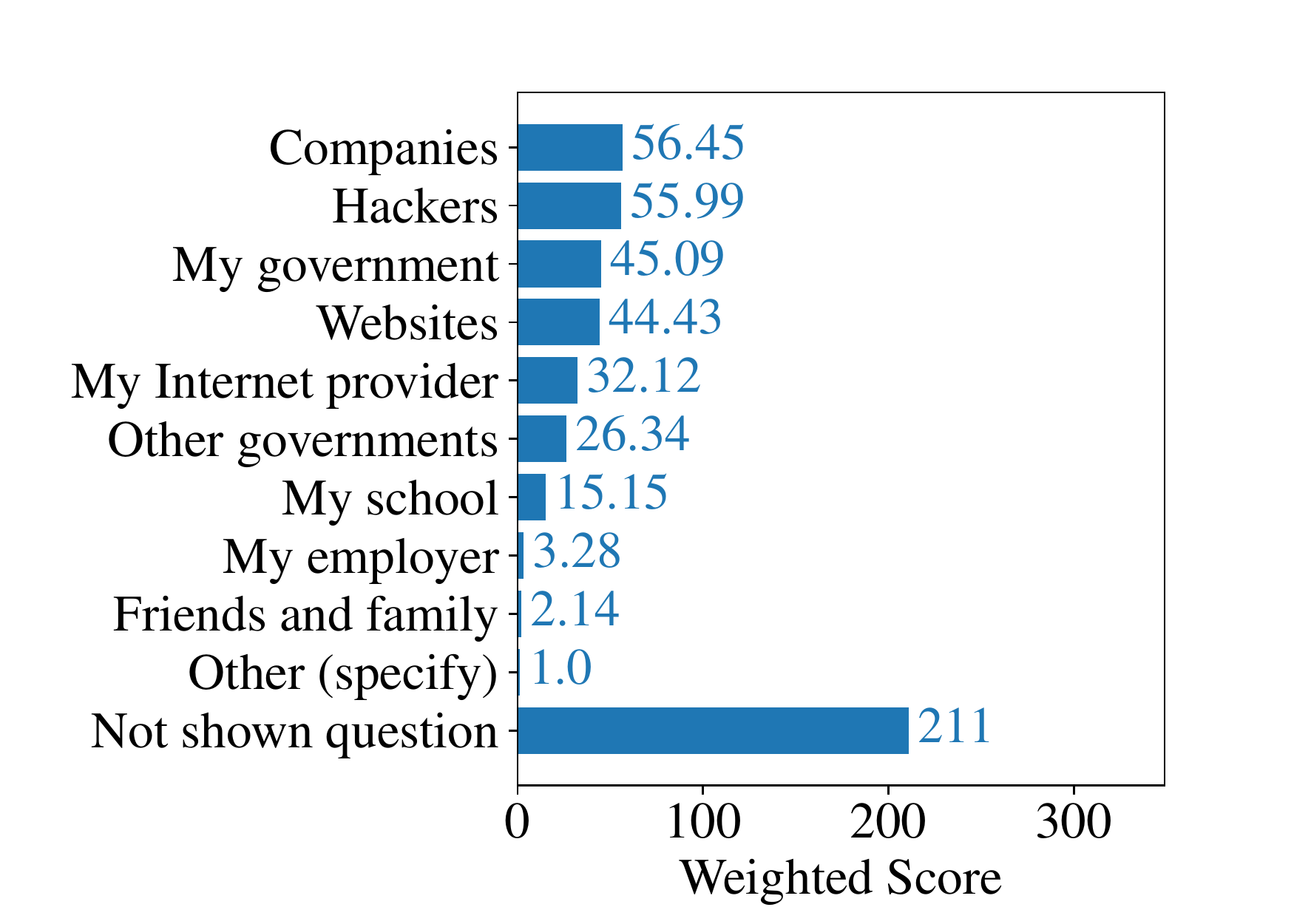}
        \caption{Who are you trying to protect\\yourself from? Choose and rank your\\choices based on level of concern.}
        \label{fig:q5_18_who_protecting_from}
    \end{minipage}
    \hfill
    \begin{minipage}[t]{0.33\linewidth}
        \centering
        \includegraphics[width=\textwidth]{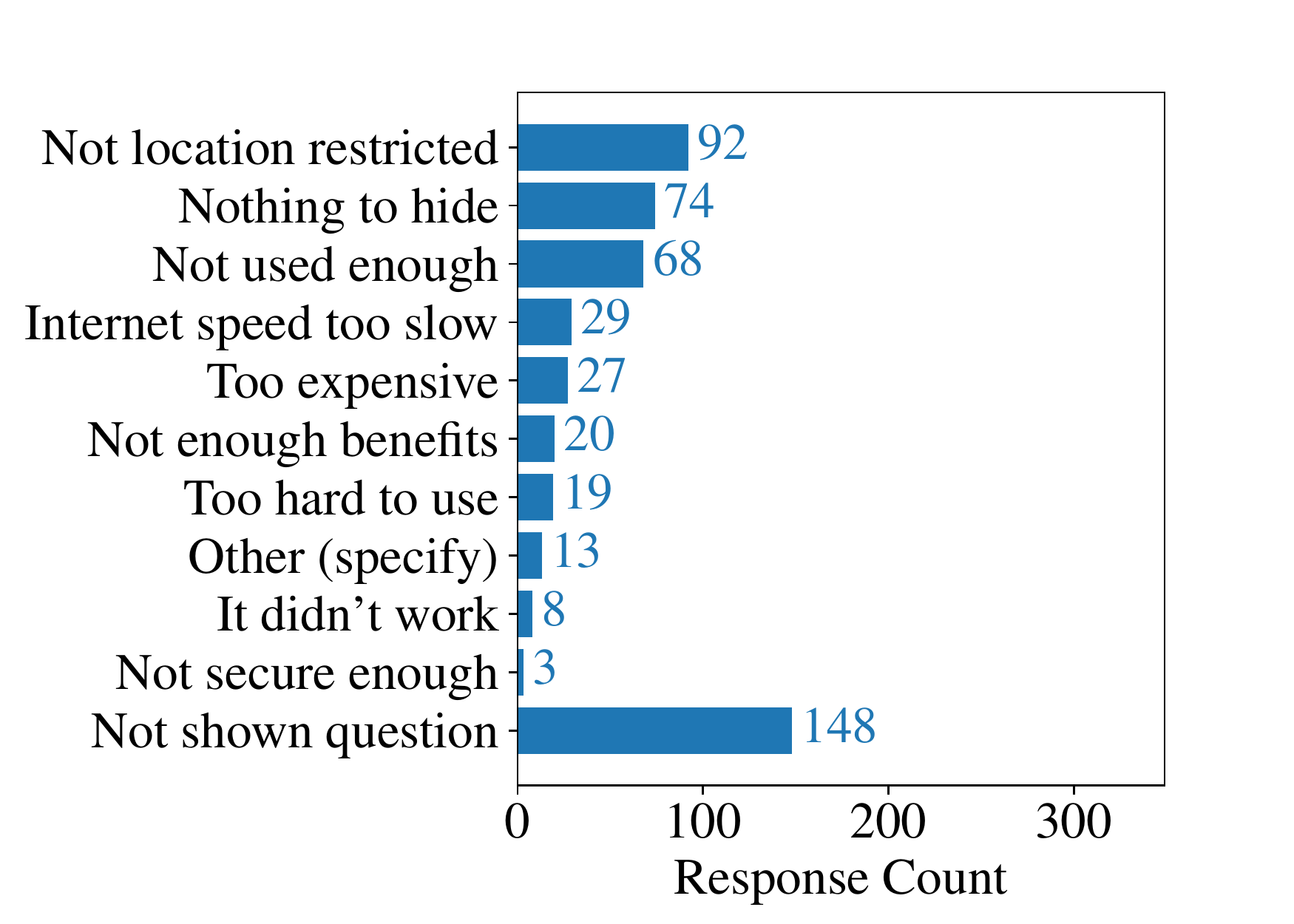}
        \caption{Why did you stop using a VPN?\\(responses selected by participants)}
        \label{fig:q5_15_stop_using_VPN}
    \end{minipage}
    \hfill
    \begin{minipage}[t]{0.33\linewidth}
        \centering
        \includegraphics[width=\textwidth]{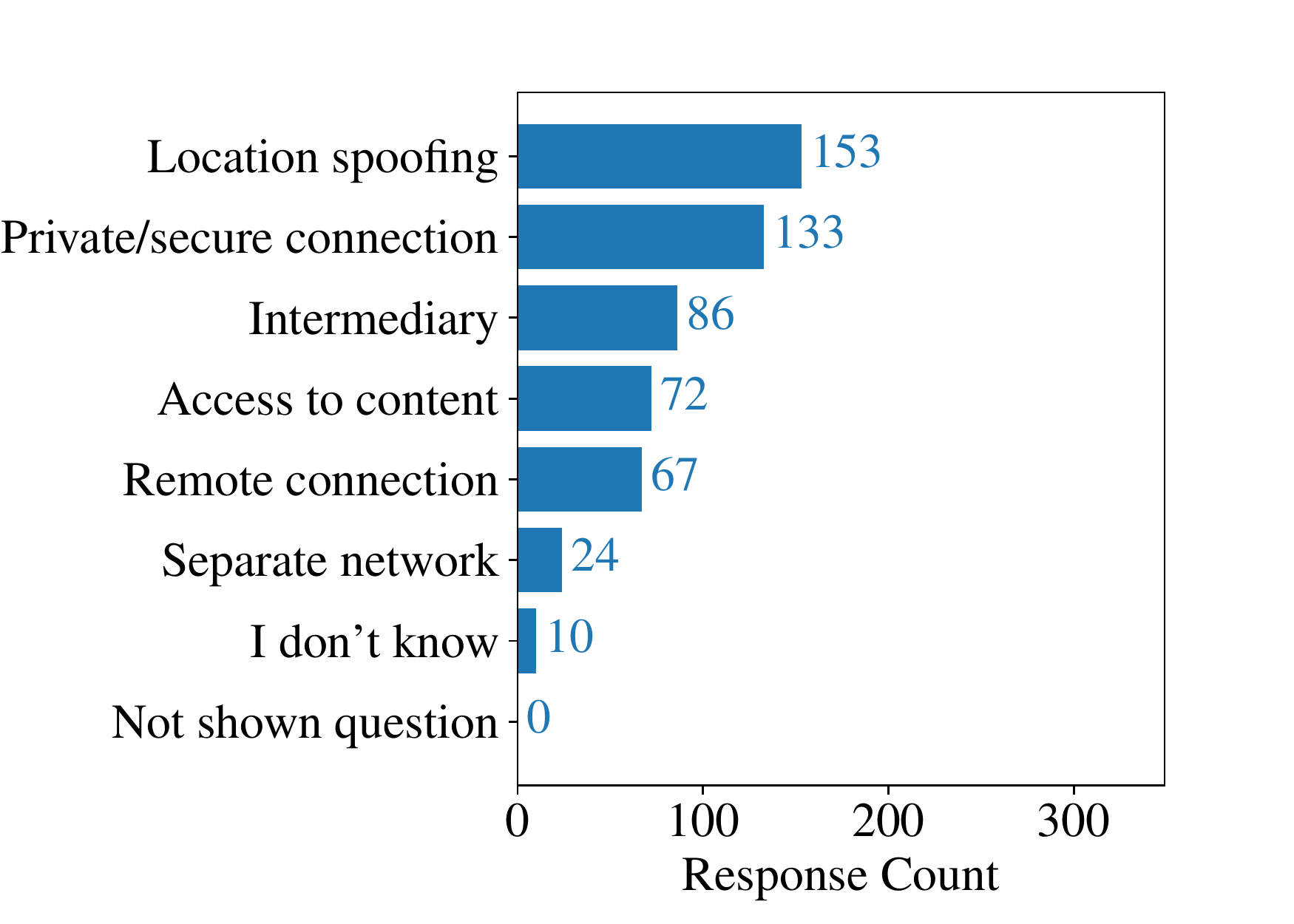}
        \caption{What do you think a VPN is?\\(response coded by researchers)}
        \label{fig:q5_1_all_participants}
    \end{minipage}
\end{figure*}

\paragraph{Survey}
In contrast to the interviews, a larger proportion of survey respondents said that they used VPNs to protect their privacy (126/349), and 105/349 said that they use VPNs for security (Figure~\ref{fig:q5_17_all_participants}). We asked these respondents to choose and rank who they were protecting
themselves from when using a VPN. Most of these survey respondents ranked
companies, hackers, the government, and websites as top concerns. 
As shown in Figure~\ref{fig:q5_18_who_protecting_from},
fewer participants were
concerned about other governments or friends and family.

For students that only used university-provided VPNs, less than a third (15/49) used VPNs to enhance privacy or 
security
(Figure~\ref{fig:q5_17_university_vpn_only}).
However, for students that only used commercial VPNs, over half (49/91) used VPNs to enhance privacy or security.
For these students, accessing blocked content was the only more common response (55/91).
As previously mentioned, for this survey 
question, there was also a significant correlation between the responses given 
by all participants and participants that only used commercial VPNs, with a 
Spearman correlation coefficient of 0.786 (p = 0.036).

\subsubsection{Students use VPNs on-demand}
\paragraph{Interview}
Seven interview participants reported using VPNs only when they needed to, while 4/32 participants would always have it on. One interviewee from the former group explained that VPNs would take up storage on their computer, as well as battery life while using it. Another example of on-demand usage was from P26, who was restricted by the bandwidth limitations of Windscribe, a free VPN: 
\begin{quote}Like the Windscribe, I get 10GB every month, and I certainly go through more than 10GB of internet.  \end{quote}

\paragraph{Survey}
VPN usage appeared to be more irregular and on an ``as needed'' basis
among both interview participants and survey respondents. Most survey respondents
(201/349) reported they did not currently use a VPN, with only 148/349 survey
respondents reporting that they currently use a VPN. When asked how often survey
respondents use a VPN, 302/349 reported only using a VPN sometimes or rarely. A
minority reported using a VPN all the time (10/349) or most of the time~(37/349). 
Figure~\ref{fig:q5_15_stop_using_VPN} shows that of the 201 respondents that stopped using VPNs, some reported that they were no longer location restricted (92/349), did not
have anything to hide (74/349), or simply did not use it enough (68/349). Very few of these respondents reported a lack of security (3/349) to be a
contributing factor in their decision to stop using VPNs.

\subsection{\mbox{Mental Models}}\label{sec:part1_mental_models}

We studied students' mental models of VPNs and
found that students could generally define a VPN but were less
familiar with how VPNs work.

\paragraph{Interview}
We found that most interview participants had a fairly good idea about the purpose of a VPN.
However, most participants were less familiar with technical explanations.
Almost half (14/32) of the interview participants described
a VPN as routing your Internet activity through third party machines or as a
service for changing your IP address, masking your identity (10/32), or
reducing others ability to track you (10/32). P18 explained:  \begin{quote}It's sort of a
    middle man. So instead of you actually downloading the file from someplace
    where somebody might be looking at you downloading it, they download it
    for you and then they send it to your computer. So it figures that they
downloaded it and not you.\end{quote} 
Some participants believed that VPNs allow you to access
blocked content (13/32), allow access into another network (7/32) and others
described a VPN as secure,  private, or adding an extra level of safety
(13/32).  In a quote typical of what we heard in participants, P25 described benefits of using a VPN: \begin{quote} Its usefulness is pragmatism, it's like, ``I need to see
    this YouTube video, but they don't let me see it in Brazil so I'm just
    going to do it in Belgium.'' I think that that's what VPNs are to me.
\end{quote}

\paragraph{Survey}
Similarly, when asked what a VPN is (Figure~\ref{fig:q5_1_all_participants}), most survey participants could list features of a VPN such
as location spoofing (153/349). As S255 described, for them a VPN was
{``Tricking my Internet to think I'm somewhere else in the world.''}
Survey respondents also described a VPN as a private or secure connection
(133/349). As S283 reported: \begin{quote}It's been described to me
as an ``Internet condom." It protects your Internet information by setting up
a different IP address.\end{quote} Other survey respondents defined a VPN as
an intermediary~(86/349), for example S78 reported that {``It's a porthole to allow
private communication/data transfer between two devices.''} Ten survey
respondents reported they did not know what VPNs are or how to define them. Overall, our participants seemed to have a good functional model of VPNs, i.e., what a VPN does for me, as opposed to a technical model.

\begin{figure}[t]
    \begin{minipage}[t]{0.49\textwidth}
        \centering
        \includegraphics[width=0.85\columnwidth]{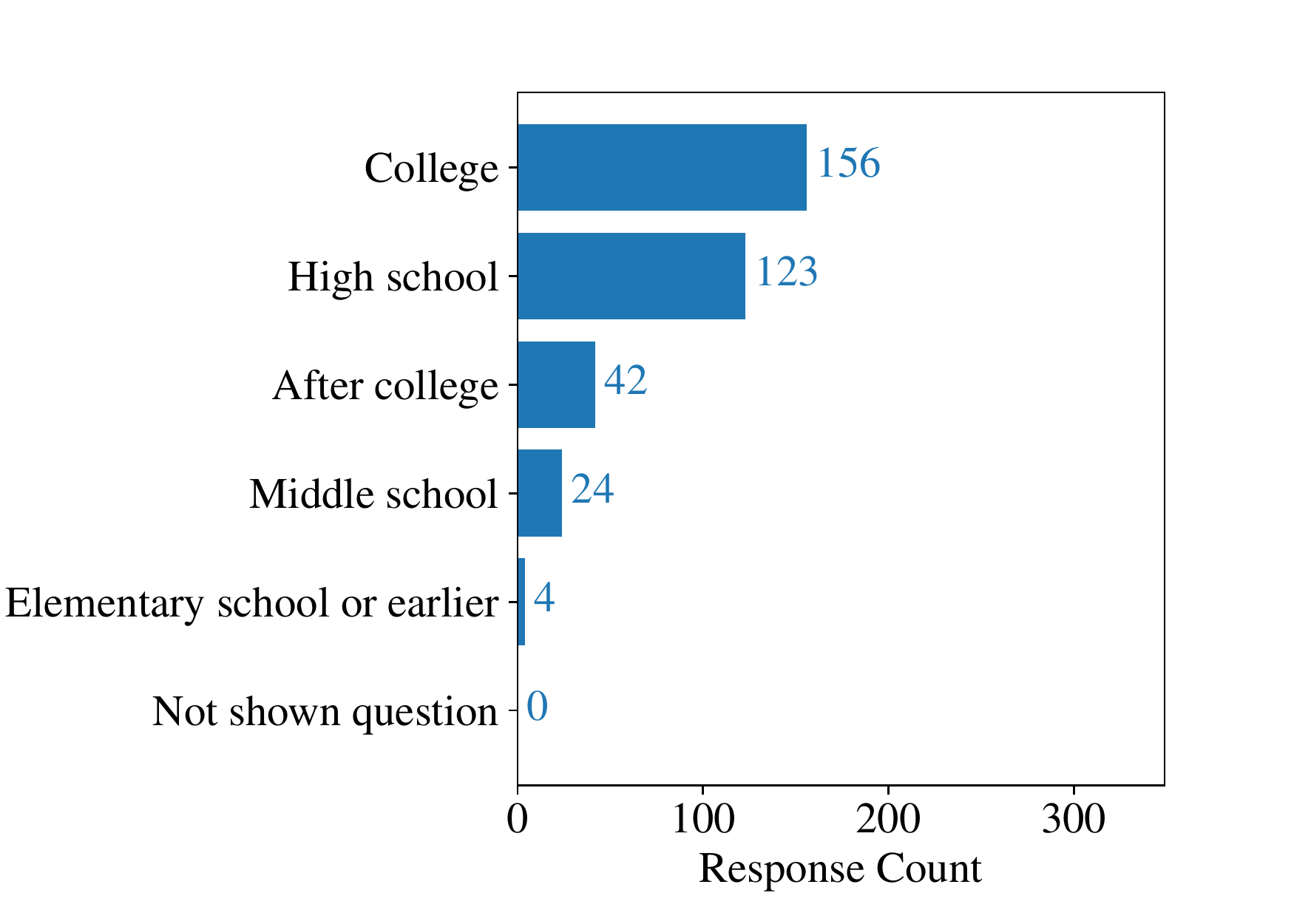}
        \caption{When did you first use a VPN? (response selected by participants)}
        \label{fig:q5_13_first_VPN_user}
    \end{minipage}
    \hfill
    \begin{minipage}[t]{0.49\textwidth}
        \centering
        \includegraphics[width=0.85\columnwidth]{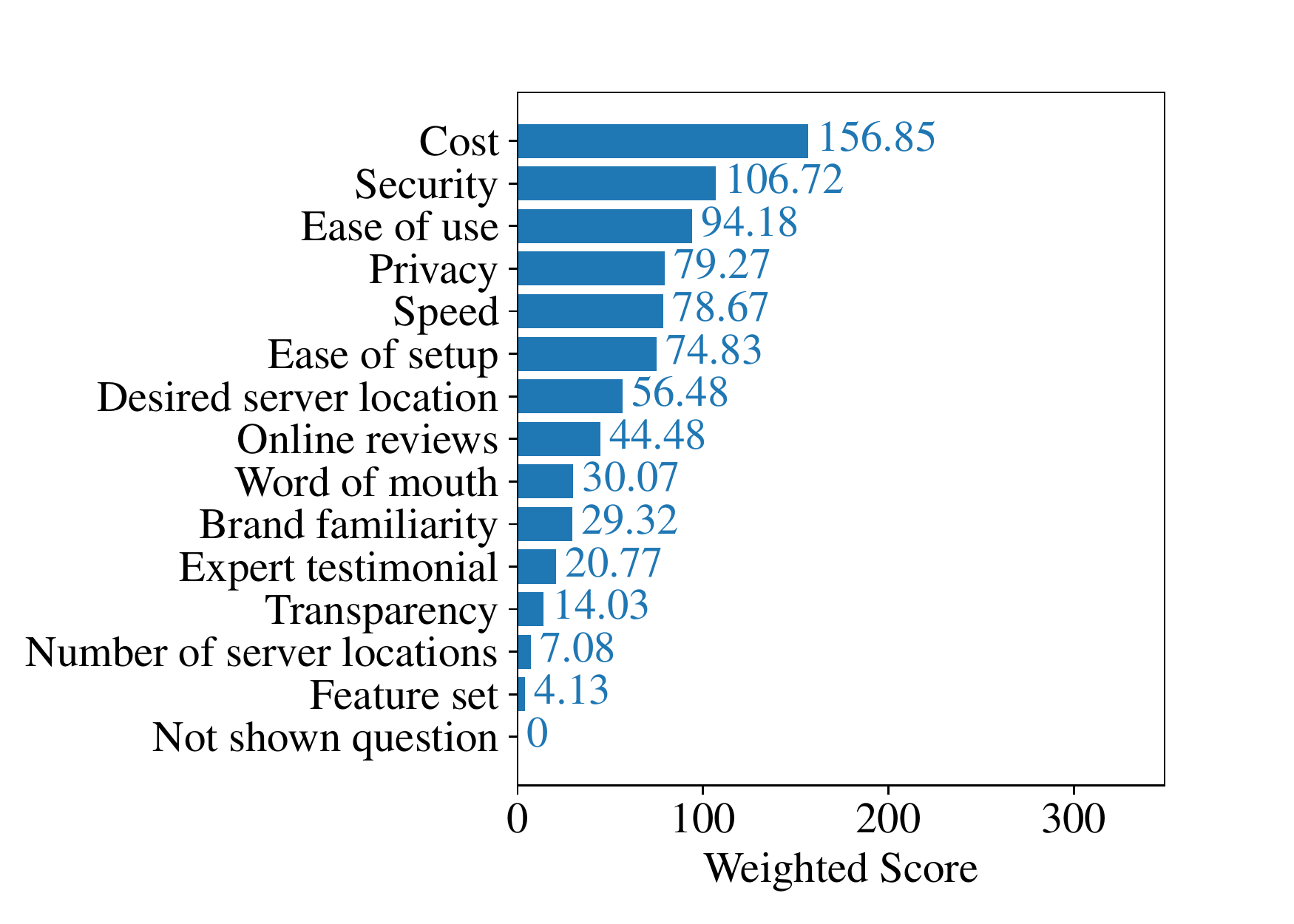}
        \caption{Choose and rank the 5 most important factors for you when choosing a VPN}
        \label{fig:q5_9_factors_choosing_VPN}
    \end{minipage}
\end{figure}

\subsection{How Students Choose VPNs} \label{sec:part1_vpn_choice_factors}

Many students began using VPNs before
entering university, and ranked cost, security, and ease of use ahead of
privacy when choosing a VPN. Most students also liked VPNs more
for accessing content rather than privacy and security.

\subsubsection{\mbox{Students learn about VPNs in high school \& college}}
\paragraph{Interview}
Two interview participants reported that they first started using VPNs when they were in high school.
Typifying what we heard, P20 told us how he used it to get access to sites that were blocked by his high school: 
\begin{quote}I've used them for a few reasons,
    but privacy was never really one of them. It was just when my content was
    restricted when I was in boarding university, I went to boarding
    university for high school. Our Wi-Fi was very tightly patrolled. So any number
    of things were blocked, like from adult content, to a lot of sports
    websites for instance were blocked, because they ``encouraged gambling'' and I like to watch a lot of sports online illegally, because that was the
    only way I could watch them.\end{quote}
    Another participant, P26, shared how they used a VPN to download a graphics editor, which they could not afford in high school: \begin{quote} The university computers came with a standard photo editor that was pretty bad. So we wanted to use Photoshop, and Photoshop is very expensive. So one of my friends recommended that we torrent it from The Pirate Bay, so we went on there, and I remember it has a warning that says, make sure your IP is masked(\dots) I did that, and then we downloaded Photoshop for a university project. I think I was maybe 16 at the time.\end{quote}

Nevertheless, some interview participants found it hard to learn about VPNs,
as P18 said: \begin{quote} I've actually never heard VPNs brought up in conversations among my friends. That's because I think they're pretty obscure at the moment. The usage of this VPN  hinges on two things, the desire to obtain copyrighted material for free and also the knowledge of the existence of VPNs. Those are two I think pretty big bottlenecks that limit this sort of information to tech nerds.\end{quote}

\paragraph{Survey}
This sentiment was confirmed by our survey respondents who reported first using a VPN at their university (156/349) or their high school 
(123/349)~(Figure~\ref{fig:q5_13_first_VPN_user}).

\begin{figure}[t]
    \begin{minipage}[t]{0.49\textwidth}
        \centering
        \includegraphics[width=0.85\columnwidth]{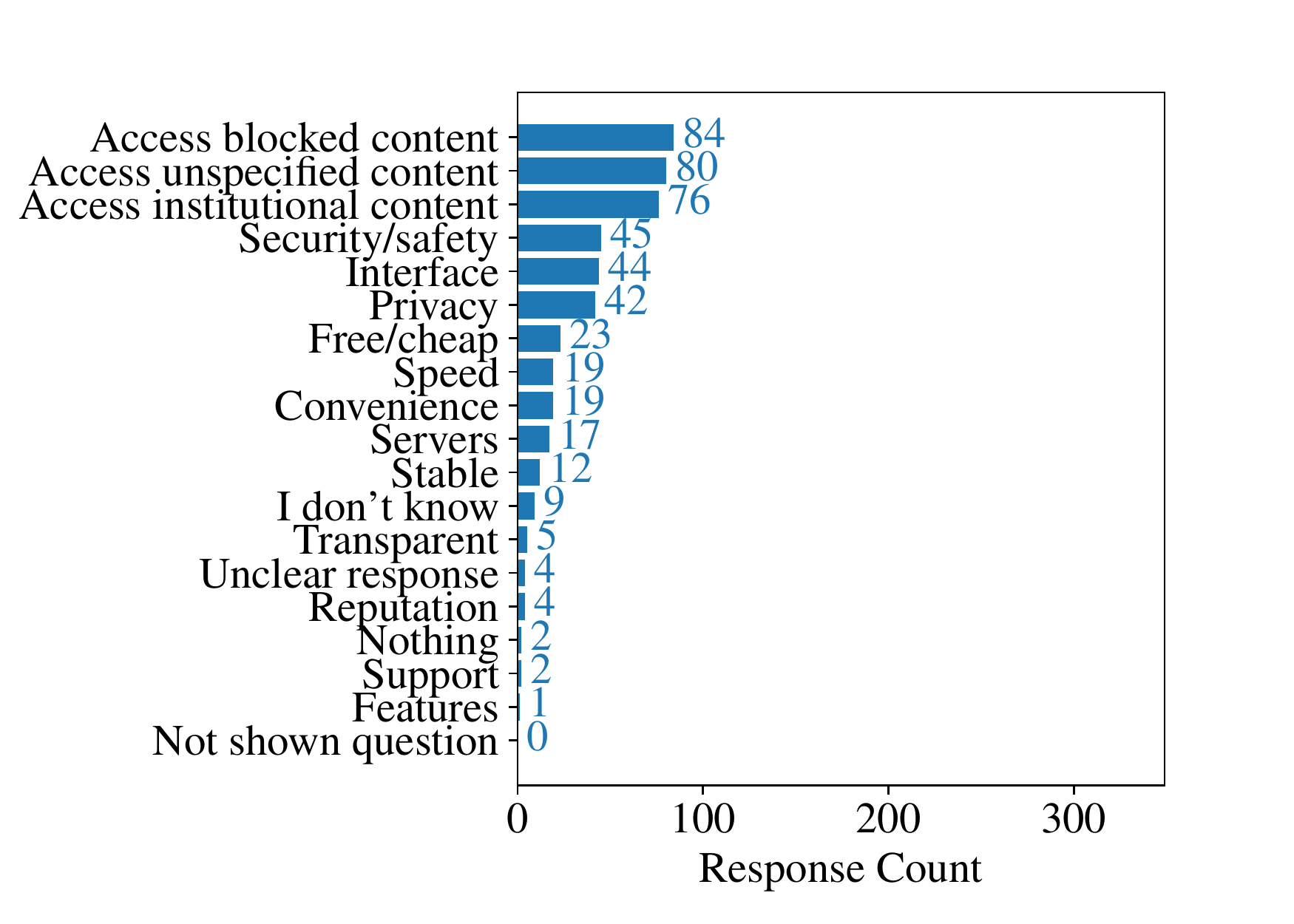}
        \caption{What do you like about your VPN(s)? (response coded by researchers)}
        \label{fig:q8_2_what_like_VPNS}
    \end{minipage}
    \hfill
    \begin{minipage}[t]{0.49\textwidth}
        \centering
        \includegraphics[width=0.85\columnwidth]{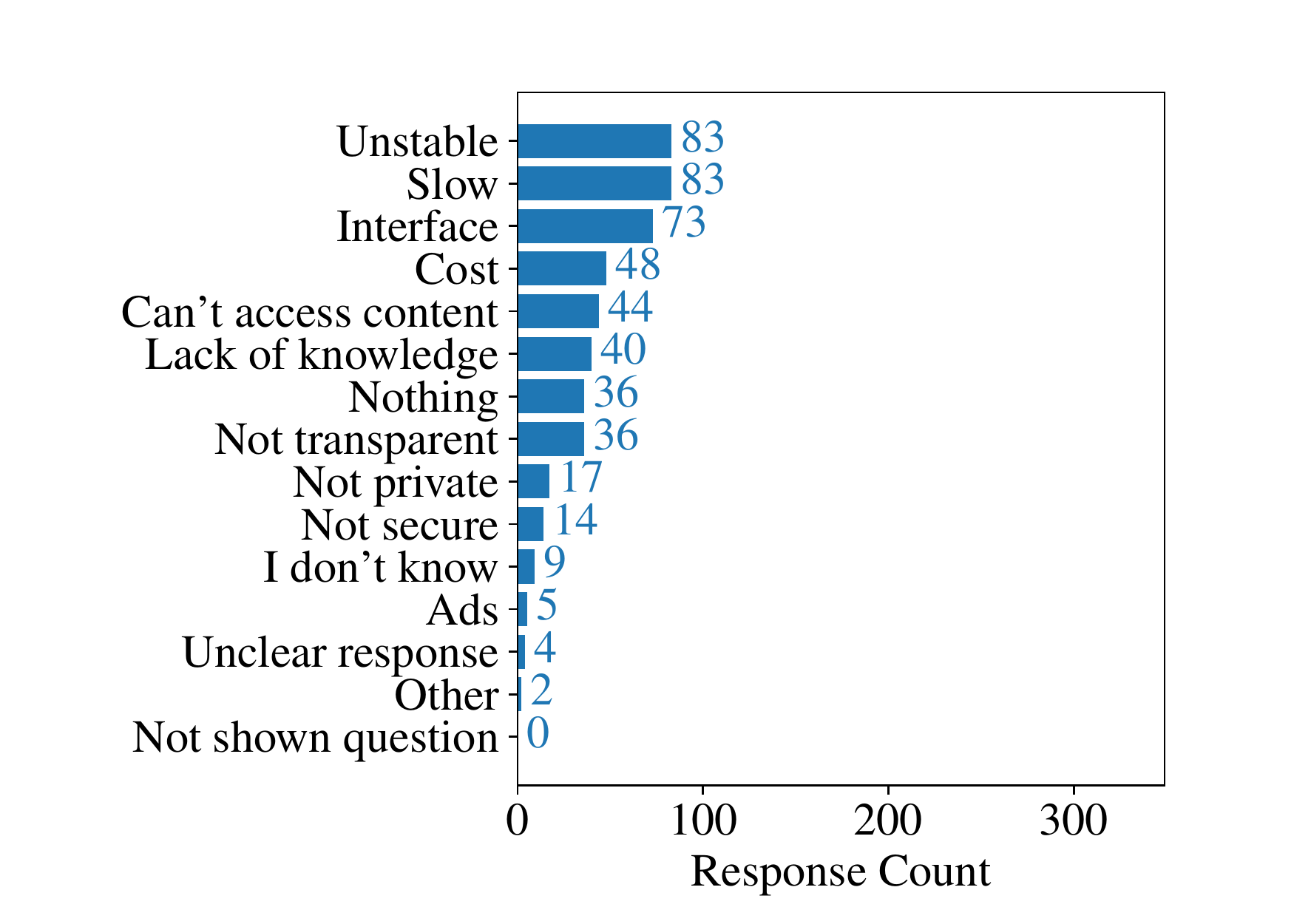}
        \caption{What do you dislike about your VPN(s)? (response coded by researchers)}
        \label{fig:q8_3_what_dislike_VPNS}
    \end{minipage}
\end{figure}

\subsubsection{\mbox{Students consider many factors to choose a VPN}} 
\paragraph{Interview}
Interviewees and survey respondents differed slightly in terms of how they chose a VPN to use. For most interview participants (19/32), the most
important factor was that the VPN provider had a good reputation; three interview participants added
that if their friends had used a VPN before, then they were more likely to use one. Ten interview participants had various
security and privacy requirements, such as making sure that the VPN had a secure network,
that the VPN provider did not store any of user' records, that the VPN provider did
not sell users' information, and that the VPN protects users' data. For one
participant it was important that a VPN did not require any personal information
when setting up the account and another one wanted an option of secure
payment. 
When we asked interview participants how they determined whether
their VPN provider was trustworthy, 13/32 said they checked that it had
good reviews online. Another 10/32 would verify that through word of mouth and 7/32
knew it was trustworthy because of who provided access to their VPN, such as the university. 

Our interview participants also indicated that  
ease of use (8/32), speed (7/32), cost (6/32), and ease of set up (5/32) were important.
Five interview participants
said that they looked at the price before purchasing a subscription; five said
that it was 
important that VPN was for free; and four said that
would always choose a cheaper option. For example, for P11, the main factors were word of mouth, experts' opinion,
cost as well as customer service available: \begin{quote} I look
    on Tech Radar and PC Monitor, those kinds of websites (\dots) I get some reviews from friends (\dots) When I came to China, I was deciding between Express and
    Astro, and I just looked on their websites, went through server
    locations, cost, and their privacy policies, (\dots) available
customer service, which was very important as well.\end{quote}

\paragraph{Survey}
On the other hand, as shown in Figure~\ref{fig:q5_9_factors_choosing_VPN}, the most important considerations as ranked by survey respondents 
for choosing between VPNs were cost, security, and ease of use. Fewer participants valued privacy, and speed and transparency were rated fairly low overall.

\begin{figure*}[t]
    \begin{minipage}[t]{0.32\textwidth}
        \centering
        \includegraphics[width=\linewidth]{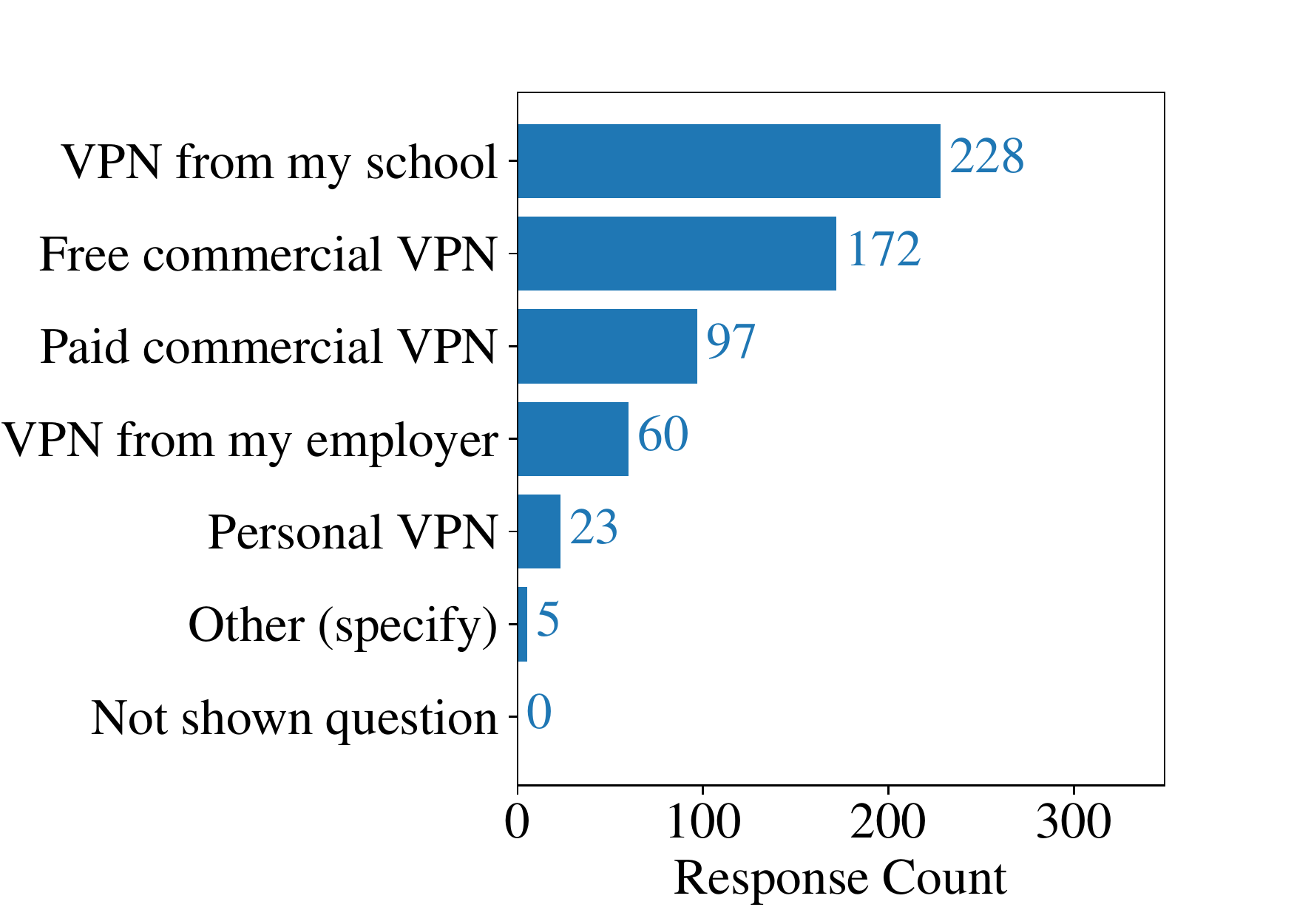}
        \caption{Types of VPNs used (responses selected by participants)}
        \label{fig:q5_3_Types_VPNS_Used}
    \end{minipage}
    \hfill
    \begin{minipage}[t]{0.32\textwidth}
        \centering
        \includegraphics[width=\linewidth]{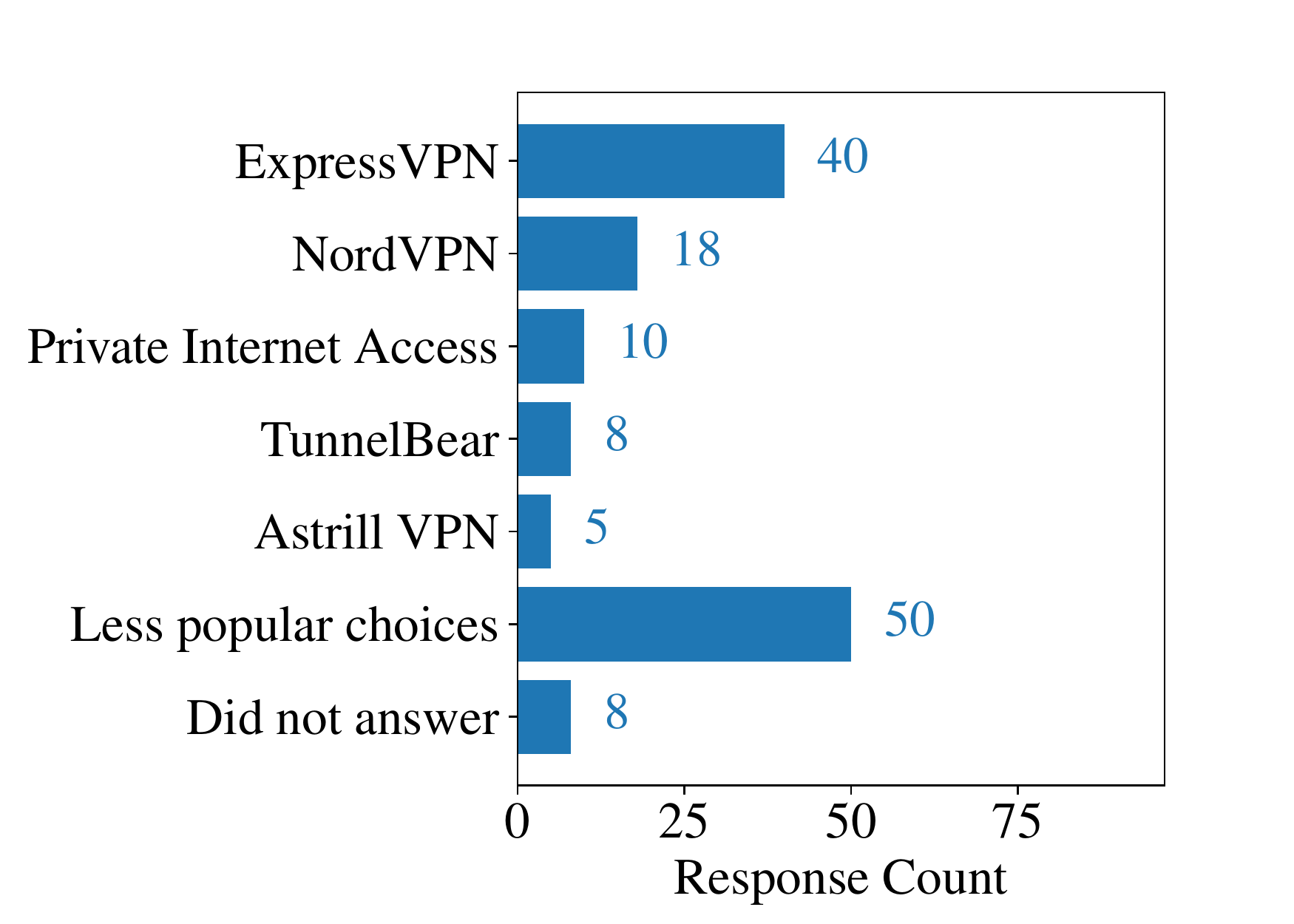}
        \caption{Popular paid commercial VPNs for the 97 respondents in Figure~\ref{fig:q5_3_Types_VPNS_Used} (responses selected by participants)}
        \label{fig:q5_4_Paid_commercial}
    \end{minipage}
    \hfill
    \begin{minipage}[t]{0.32\textwidth}
        \centering
        \includegraphics[width=\linewidth]{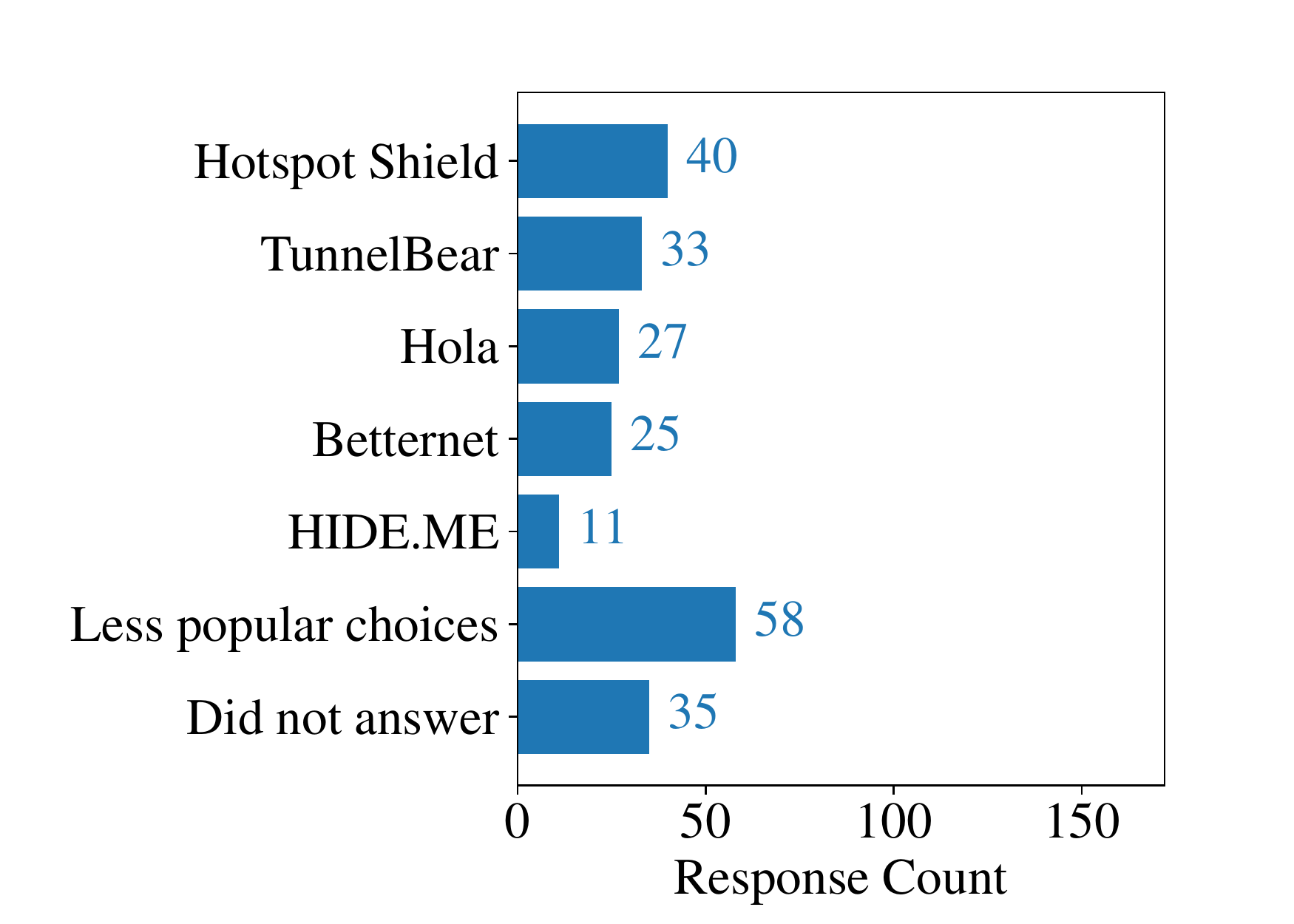}
        \caption{Popular free commercial VPNs for the 172 respondents in Figure~\ref{fig:q5_3_Types_VPNS_Used} (responses selected by participants)}
        \label{fig:q5_5_Free_commercial}
    \end{minipage}
\end{figure*}

\subsubsection{Students value access over security \& privacy}
\paragraph{Interview} We asked interviewees about their general experience and
feelings related to VPN usage. We asked whether they saw any differences
in their Internet experience when using VPNs. Ten interviewees found the
biggest difference in being able to access blocked content, and 8/32 felt more
secure. Nevertheless, 10/32 interviewees did not see any difference in the way
they used the Internet and did not observe changes in their online habits. P20
noted: \begin{quote}I sort of have the assumption that any time I use the
Internet, any privacy I have is super limited. But you would think that using
a VPN would help with that in some way. I don't think it would actually change
my behavior online at all, but I think it would definitely make you feel a bit
more secure in that.\end{quote}

\paragraph{Survey} We asked survey respondents to report, in short-answer
form, what they liked and disliked about VPNs, as shown in Figure
\ref{fig:q8_2_what_like_VPNS} and Figure \ref{fig:q8_3_what_dislike_VPNS}.
The ability to access blocked content (e.g. geo-restricted video streaming
websites), institutional content, or other kinds of content was by far the
most commonly liked feature of students’ VPNs (223/349). S131 appreciated that
they could have access to many things:\begin{quote}It allows me to view
    content that is restricted by a time zone limit like test scores,
    acceptance letters etc. Also if you're in another country that doesn't
allow certain media platforms (e.g., Netflix, Hulu), VPNs allow you to access
them.\end{quote} Other qualities, including security, privacy and interface
received far fewer mentions.  Survey respondents did not like slow (83/349)
and unstable (83/349) VPN connections. S192 shared their frustration:
\begin{quote}The connection is very slow and unstable, and it would often turn
    off while I was accessing the Internet so that I would have to reload
everything.\end{quote} Survey respondents did not like the complexity of
interfaces and their features. As S306 explained: \begin{quote}It is annoying
    to log into all the time, and it automatically logs me out after a
    designated amount of time.\end{quote} and S279: \begin{quote}It is
difficult to use and the set-up process is confusing. Selecting a new VPN is
very confusing because there are so many options.\end{quote} Forty-eight
survey participants also complained about cost. For example, they did not like
that free VPNs have limited server locations and they have to pay for
unlocking more. 

\begin{figure}[t] \centering
    \includegraphics[width=0.85\columnwidth]{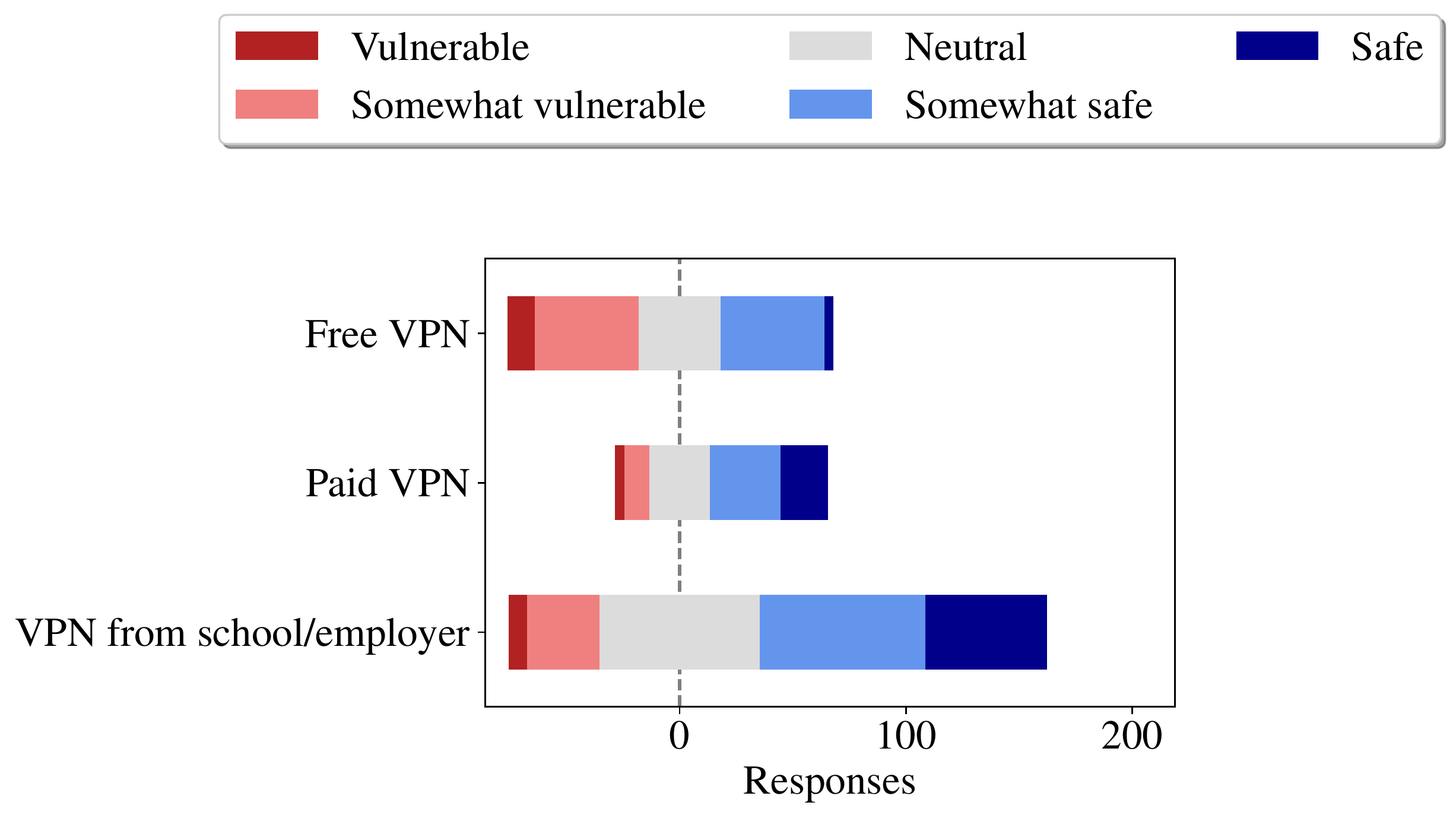}
    \caption{How ``safe'' participants feel using different types of VPNs.
    (responses selected by participants)}
    \label{fig:q5_6_Safe_Using_Free_commercial} \end{figure}

\subsection{Which VPNs Students Choose} Students study felt most comfortable
with using VPNs provided by their institutions. They were puzzled whether
commercial VPNs actually provide privacy and security.  Thus, they believed
that a university VPN was a safer choice.

\subsubsection{University VPNs are most prevalent} \paragraph{Interview}
During the interviews, we learned that 16/32 of participants used work related
or institutional VPNs. Among them, 12/16 specifically reported using VPNs
provided by their university, and 6/16 had never used any commercial VPNs. We
also found that 10/32 interviewees used free VPNs and 7/32 used paid or free
trial version of a paid VPN. 

\paragraph{Survey} We observed a similar pattern in the responses from survey
participants. Most survey respondents used the VPN offered by their university
(228/349). Nearly half also used free commercial VPNs (172/349). A smaller
fraction of students study used paid commercial VPNs (97/349).
Figure~\ref{fig:q5_3_Types_VPNS_Used} summarizes these results.  Fewer survey
respondents used VPNs through their employer (60/349) or a personal VPN that
they set up themselves (23/349).

Figure~\ref{fig:q5_4_Paid_commercial} shows the most common paid commercial
VPNs that students we surveyed reportedly used. We find that survey
respondents used a variety of paid VPNs, including ExpressVPN (40/97), NordVPN
(18/97), and PrivateInternetAccess (10/97).  There is also a long tail of less
popular choices, with 50 students using paid commercial VPNs that five or less
other students used.  Eight survey participants that reportedly used paid
commercial VPNs did not specify particular VPNs.

Figure~\ref{fig:q5_5_Free_commercial} shows the most common free commercial
VPNs students we surveyed reportedly used: Hotspot Shield (40/172), TunnelBear
(33/172), Hola (27/172), and Betternet (25/172).  Notably, some respondents
indicated they used SonicWall or ConnectTunnel which is the VPN offered by the
university, indicating some confusion on what is an institutional versus
commercial VPN provider.  Furthermore, 35 surveyees that reportedly used free
commercial VPNs did not specify which VPNs they used.

\subsubsection{Students feel safer using their university VPN} Students were
more comfortable with university and paid commercial VPNs than free commercial
VPNs. 

\begin{figure}[t!] \centering \begin{subfigure}[t]{0.49\textwidth}
\includegraphics[width=\linewidth]{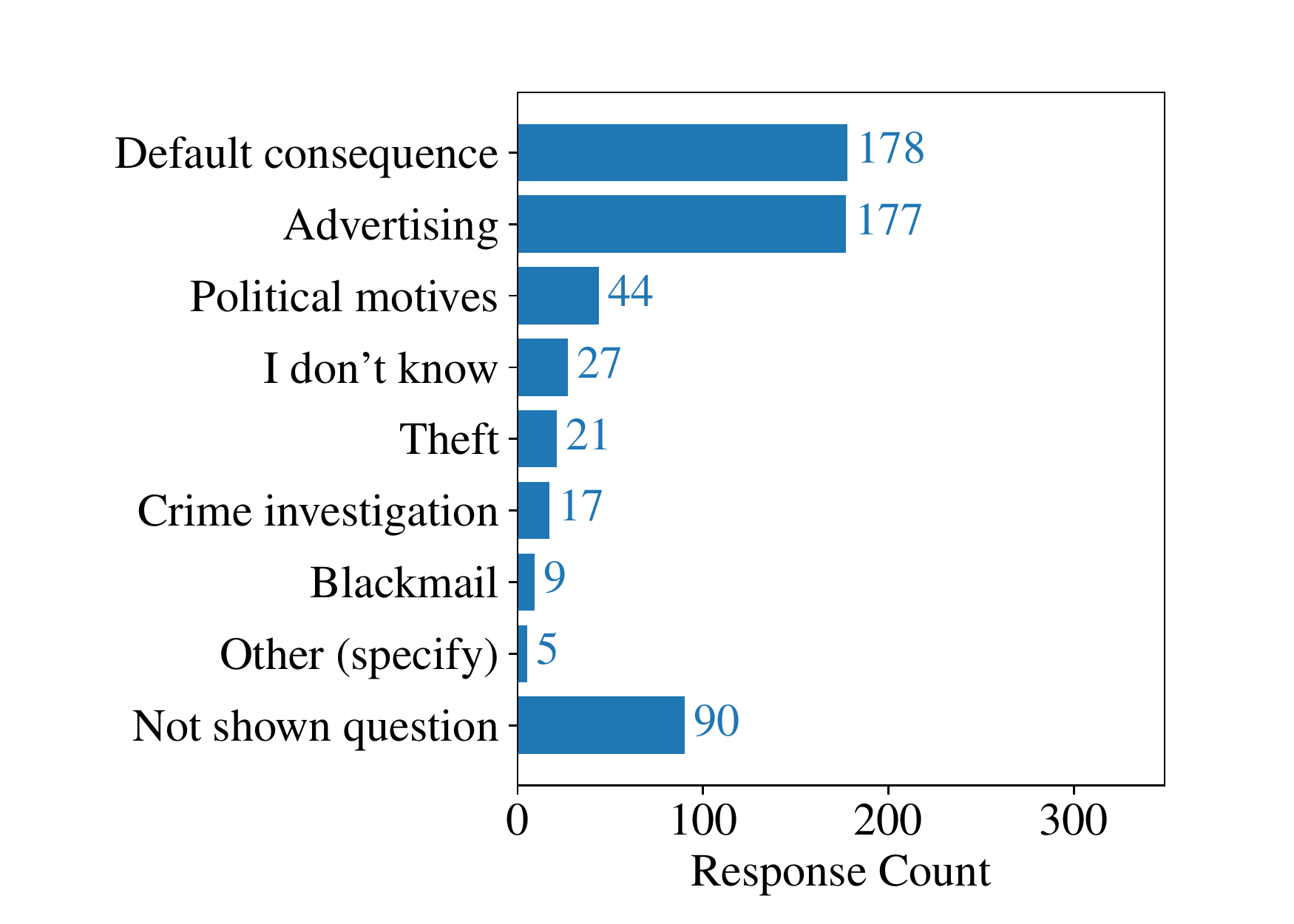} \caption{All
participants.} \label{fig:q7_6_all_participants} \end{subfigure} \hfill
\begin{subfigure}[t]{0.49\textwidth}
\includegraphics[width=\linewidth]{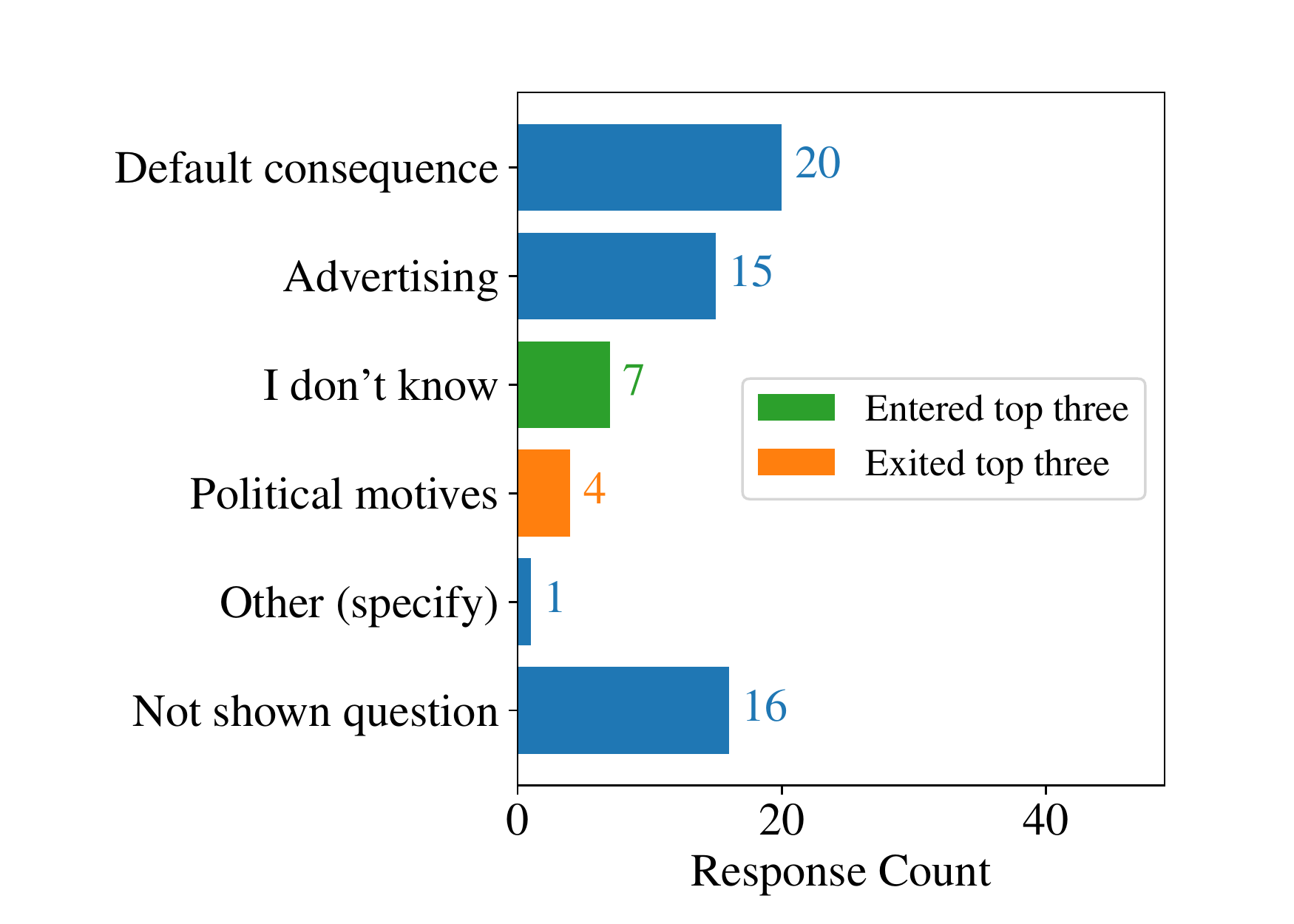}
\caption{Participants that only used university VPNs.}
\label{fig:q7_6_university_vpn_only} \end{subfigure} \caption{Why do you think
your VPN provider collects your data? (responses selected by participants)}
\label{fig:q7_6_why_VPN_collects_data} \end{figure}

\paragraph{Interview} When asked whether it was important who their VPN
provider was, 11/32 interview participants reported that it was important,
especially for these who used their university's VPN (7/11). University VPNs
were reassuring for them because they believed they were safe to use. However,
interviewees were split on what they were willing to do online using their
university or work related VPN. For instance, 5/16 interview participants who
used their university VPN reported that they would use it only for completing
university work, because they simply did not feel private, they felt that
university could track them, or  that using their university network made them
more vulnerable.  P13 gave an example about a student who shut down exams by
hacking into their university's VPN through another VPN to show how the
university VPN was still vulnerable. On the other hand, 5/16 other interview
participants told us they used their university VPN for private activities,
such as browsing. For example, P32 would simply forget to switch it off and
did not mind having it on: \begin{quote}It really doesn't bother me if someone
is looking at what I'm doing while I'm on the VPN, just because my philosophy
is like, at this point it's probably all there anyway.\end{quote} We found
that 14/32 interview participants would not pay attention to or care about who
their VPN provider was. For one of these interview participants, it was not
important that the university was their provider, as they explained--the
university is only a client of another provider, not a provider itself. 

\paragraph{Survey} Figure~\ref{fig:q5_6_Safe_Using_Free_commercial} shows that
the interview results were mirrored in that survey respondents felt safer
using university or employer provided VPNs than free commercial VPNs. 

\subsubsection{Students are confused about the privacy guarantees of both paid
and free VPNs} \paragraph{Interview} We found that 22/32 interview
participants would use free VPNs while ensuring they were safe, and 9/32 said
that they would not use them because they did not feel safe.  Nevertheless,
many were confused about the benefits of using a free versus paid VPN, as
expressed by P24: \begin{quote} I think the one that you have to pay for is
    more trustworthy.  But, it could easily be the other way around. Just
    because you have to pay for something doesn't mean that it is more
    reliable, or even more efficient. But, I do think that the paid ones
    generally people might think that they are more safe to use. And that
their information may be more secured, just because of that added price tag on
it.\end{quote} \paragraph{Survey}

Figure~\ref{fig:q5_6_Safe_Using_Free_commercial} shows how safe survey
participants felt using free VPNs, paid VPNs, and VPNs provided through their
university or employer.  We only show data for students that have used one or
more of these VPN types.  We find that, proportionally, survey participants
felt most safe or somewhat safe using paid VPNs (52/97) and
university/employer VPNs (127/238) than free VPNs (50/172).  Many more survey
participants reported using free VPNs than paid VPNs, even though they felt
less safe.  Eight participants that used paid VPNs and 35 participants that
used free VPNs did not indicate how safe they felt.

\subsection{Expectations About VPN Data
Collection}\label{sec:part1_expectations}

Students were unsure about the data collection practices of VPN providers.
They believed their data could be collected, but they did not necessarily
understand the consequences or who else could access their data.

\subsubsection{Students believe VPNs collect data about them}
\paragraph{Interview} Most interview participants (23/32) believed that VPNs
collect their data, with some expressing that VPNs keep data for statistics or
to sell the data. For example, P11 believed that VPNs could keep logs for many
different reasons:\begin{quote}If you're using VPNs for a bit more nefarious
means, for example, like buying drugs or trading child pornography and things
like that. (\dots) I think some of them do keep logs, and they're able to give
them over to police, and the governments, and things like that. (\dots) And
then, other ones are a bit more simple, like tracking people's web habits to
sell to advertisers and things like that.\end{quote} Several interview
participants (7/32) also believed that the university has a VPN to access all
information about students and to monitor if someone is cheating during exams.
\begin{figure}[t!] \centering \begin{subfigure}[t]{0.49\textwidth}
\includegraphics[width=\linewidth]{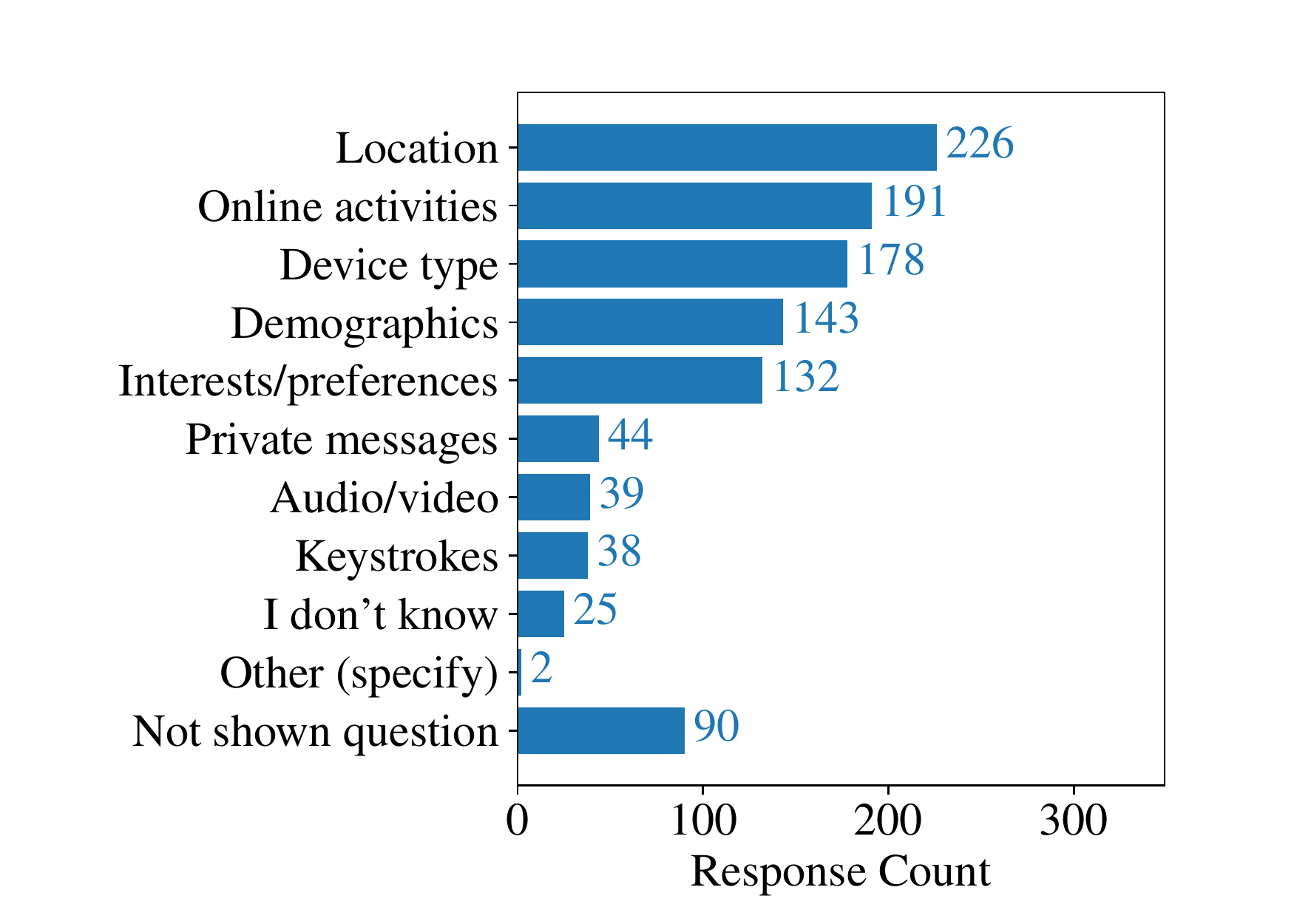} \caption{All
participants.} \label{fig:q7_5_all_participants} \end{subfigure} \hfill
\begin{subfigure}[t]{0.49\textwidth}
\includegraphics[width=\linewidth]{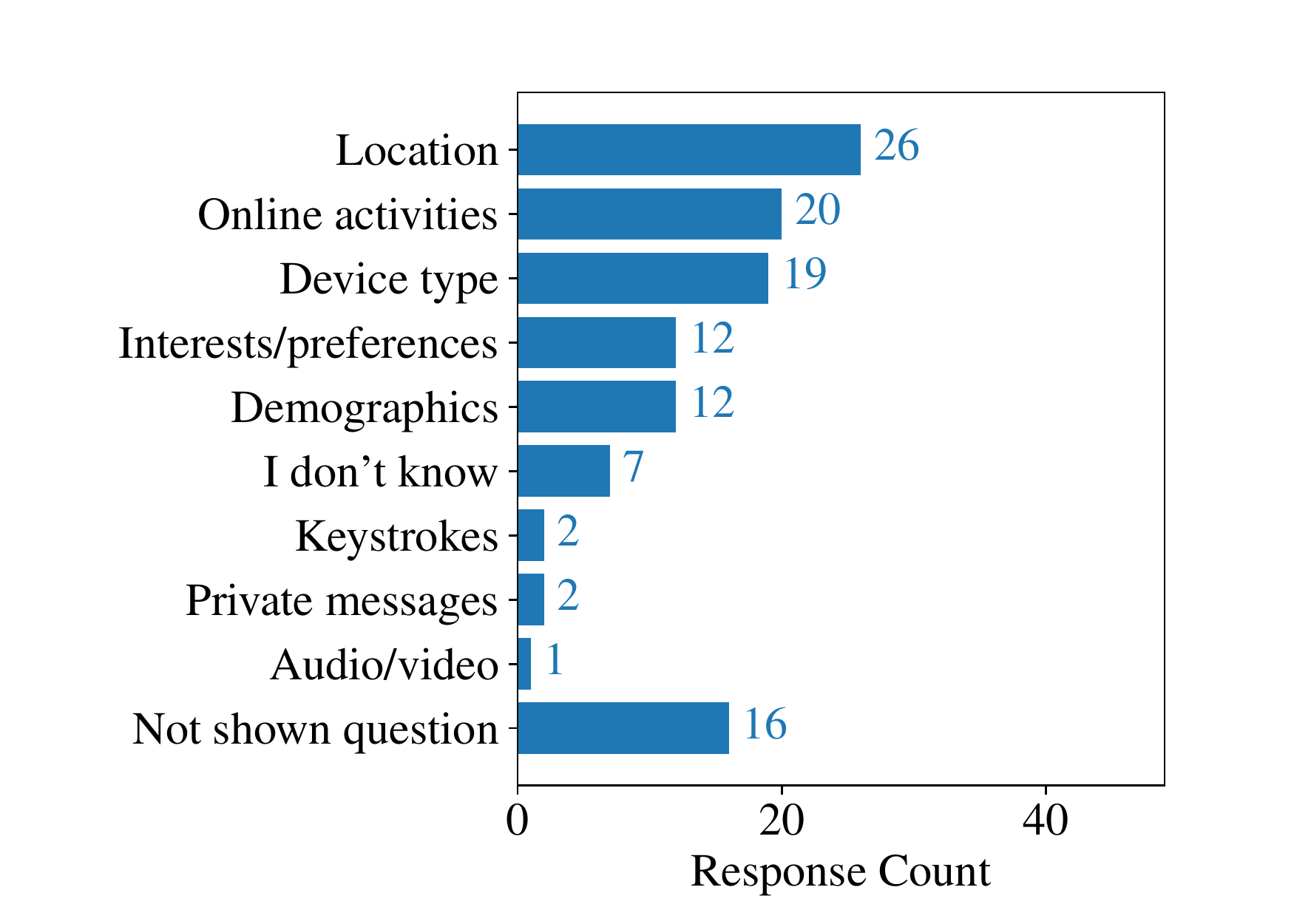}
\caption{Participants that only used university VPNs.}
\label{fig:q7_5_university_vpn_only} \end{subfigure} \caption{What kind of
data do you think your VPN provider collects about you? (responses selected by
participants)} \label{fig:q7_5_what_VPN_collects_about_you} \end{figure}

\begin{figure}[t!] \centering \begin{subfigure}[t]{0.49\textwidth}
\includegraphics[width=\linewidth]{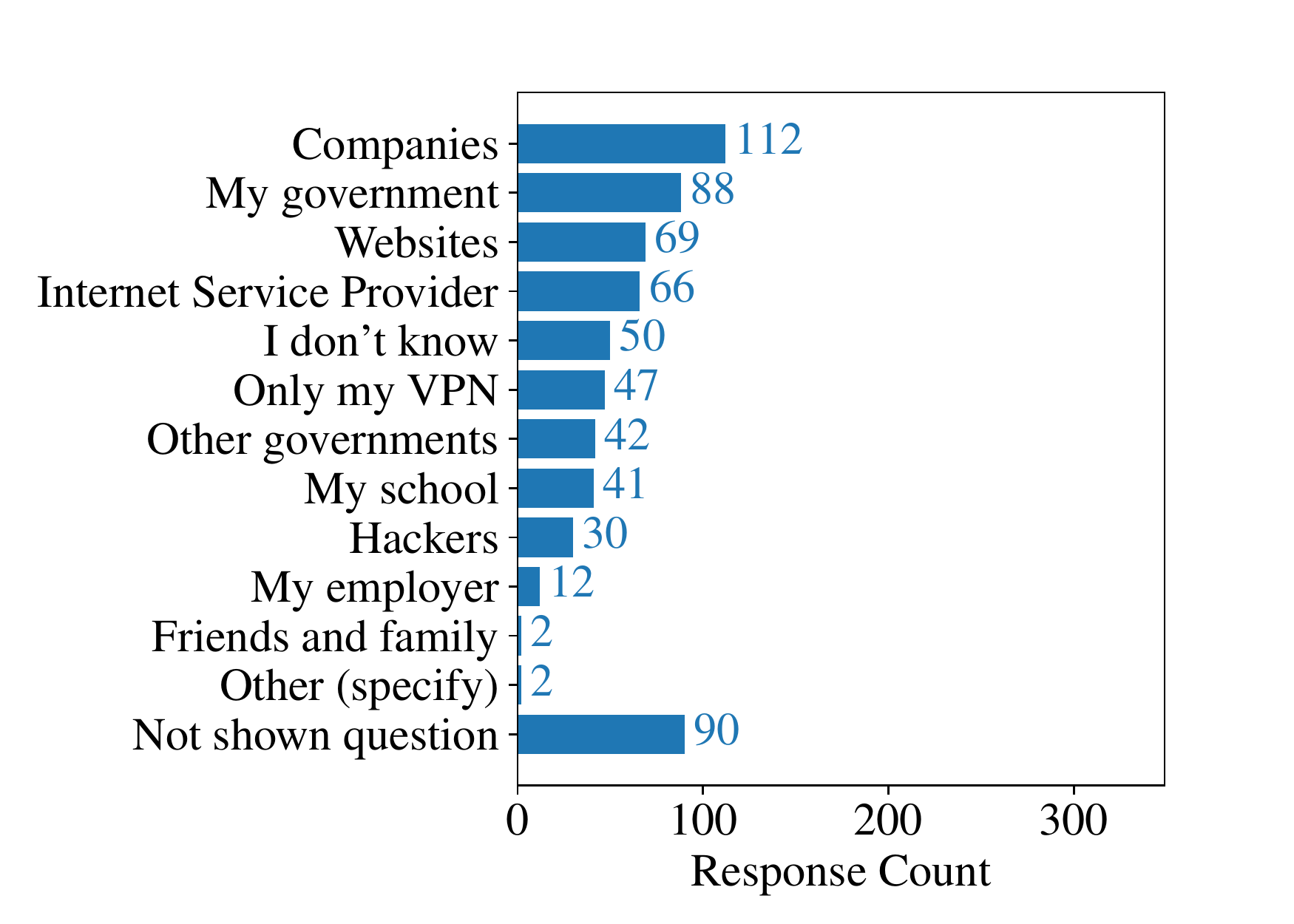} \caption{All
participants.} \label{fig:q7_7_all_participants} \end{subfigure} \hfill
\begin{subfigure}[t]{0.49\textwidth}
\includegraphics[width=\linewidth]{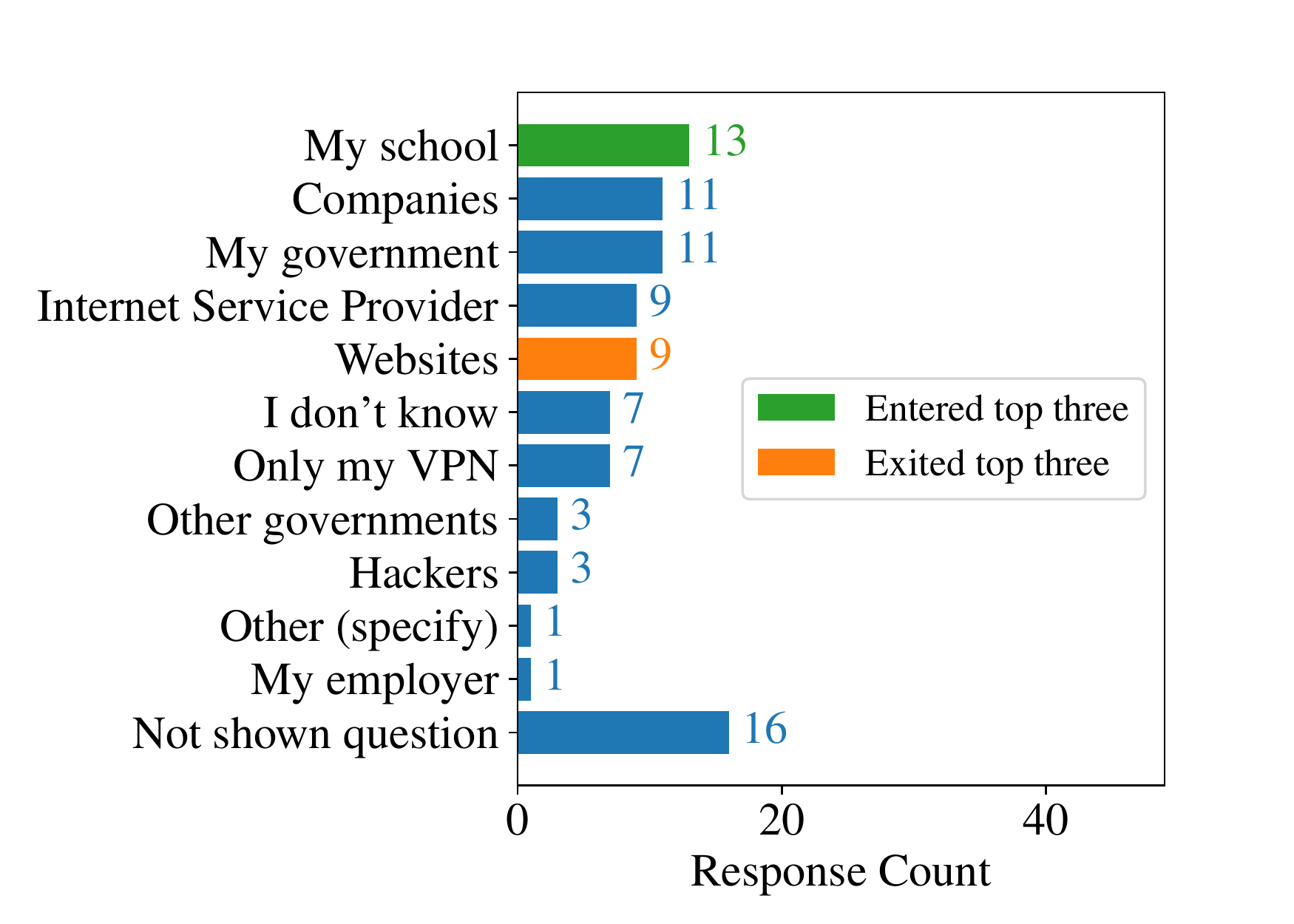}
\caption{Participants that only used university VPNs.}
\label{fig:q7_7_university_vpn_only} \end{subfigure} \caption{Who do you think
has access to the data collected by your VPN? (responses selected by
participants)} \label{fig:q7_7_who_has_access_to_VPN_data} \end{figure}

\paragraph{Survey} We found that 259/349 of survey participants similarly
believed that their VPN provider could collect data about them.
Figure~\ref{fig:q7_6_all_participants} shows that most of these 259
respondents believed that VPNs collect data for commercial motives (177/259),
or simply because data collection is a {``default consequence of using the
Internet''} (178/259); 126/259 of respondents selected both
options. Few survey respondents believed that the motives for data collection
were nefarious---such as blackmail (9/259)---and some survey
respondents selected ``I don't know'' (27/259).

We found that 33 out of the 49 survey participants who only use VPNs
provided by their university believed that their VPN provider collect data
about them.  As with the other survey participants, most of them believed that
this was both a default consequence of the Internet (20/49) and for
advertising purposes (15/49) (Figure \ref{fig:q7_6_university_vpn_only}).
Similarly, 67 out of the 91 survey participants that only used a commercial
VPN believed their VPN provider collects data about them.  Most of them
believed that this data collection was for advertising purposes (48/91), and
also a default consequence of the Internet (40/91).  For this survey question,
there was a significant correlation between the responses given by all
participants and participants that only used commercial VPNs, with a Spearman
correlation coefficient of 0.976 (p $\ll$ 0.001).

Thus, regardless of who provides their VPN or why they use it, students
believed they were being tracked.

\subsubsection{Awareness of collection, but not access \& sharing}

\paragraph{Interview} We asked interview participants about their opinion on
VPN data sharing practices.  When asked whether they thought their VPN
providers could be sharing their information, 17/32 responded ``no'' and 11/32
``yes'', but 12/32 were uncertain about their response, because they did not
feel like they would be able to know anyway. P26 explained: \begin{quote} If
they share it with someone, then they're not sharing it in a way that I would
be able to tell, because, for example, I've never seen personalized ads from
things that I've looked at while on the VPN.\end{quote} From the interview
participants who said that VPN providers do not share information with other
entities, eight confessed that they hope their information was not being
shared, and five admitted that while their VPN providers do not share any
information, other VPN providers may do so. Two of these participants believed
that even though their VPN providers do not share data with others on regular
basis, they would with legal authorities. For instance, P14 shared:
\begin{quote} If the Chinese government were to really threaten them. This is
a very hypothetical situation. I think information which they could be able to
collect and which would be interesting would probably be something like on a
service level, the actual content that you've been accessing. Like the actual
service or addresses which you've accessed recently as well as the associated
file or data which is generated while somebody is accessing the Internet.
\end{quote}

\paragraph{Survey} We asked survey participants who believed that VPNs
collected their data what they think is collected.
Figure~\ref{fig:q7_5_all_participants} shows that most participants  believed
that VPNs collect location data (226/259) and online activities~(191/259).
Fewer believed that VPNs collected private messages (44/259), recordings
(39/259), or keystrokes (38/259). Some participants did not know what was
collected (25/259).  There was also little consensus between participants on
who had access to the collected data.  The largest proportion of survey
respondents, as shown in Figure~\ref{fig:q7_7_all_participants}, believed that
companies (112/259) and the government (88/259) had access to the data
collected by VPNs. A smaller proportion believed that only VPNs had access
(47/259), and 50/259 of respondents did not know where their data went.

A plurality of survey participants who only used
university VPNs believed their VPN provider could see their location and
online activities (Figure~\ref{fig:q7_5_university_vpn_only}).  They 
believed this data was accessible by their university
(Figure~\ref{fig:q7_7_university_vpn_only}).  This result is particularly
interesting because more of these participants used VPNs to access
institutional material than to enhance privacy or to access
location-restricted content (Figure~\ref{fig:q5_17_university_vpn_only}).
Thus, it seems that even students who only used VPNs provided by their
university had a somewhat defeatist attitude about their ability to remain
private.  Similarly, a plurality of participants that only used
commercial VPNs believed that their VPN provider could see their location
(57/91) and online activities (46/91).  For this survey question, there was a
significant correlation between the responses given by all participants and
participants that only used commercial VPNs, with a Spearman correlation
coefficient of 0.997 (p $\ll$ 0.001).
Interestingly, none of these survey participants believed their university had 
access to this data. Instead, a plurality of these participants
believed that other companies had access to this data (30/91).  
We observed a significant correlation between the responses
given by all participants and participants that only used commercial VPNs,
with a Spearman correlation coefficient of 0.933 (p $\ll$ 0.001).

\begin{figure}[t] \centering \begin{subfigure}[t]{0.49\textwidth}
\includegraphics[width=0.85\columnwidth]{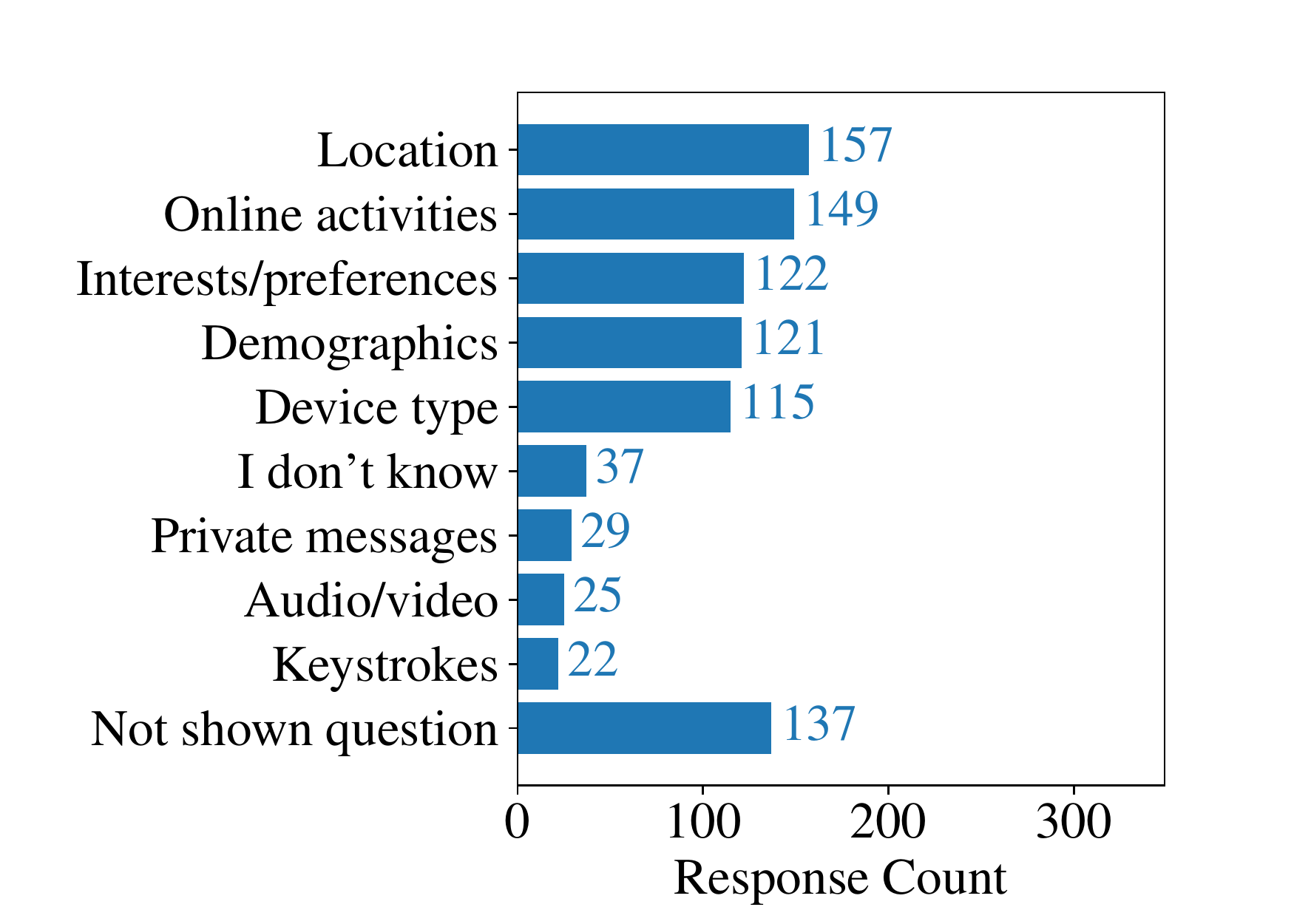}
\caption{All participants.} \label{fig:data_shared_all} \end{subfigure} \hfill
\begin{subfigure}[t]{0.49\textwidth}
\includegraphics[width=0.85\columnwidth]{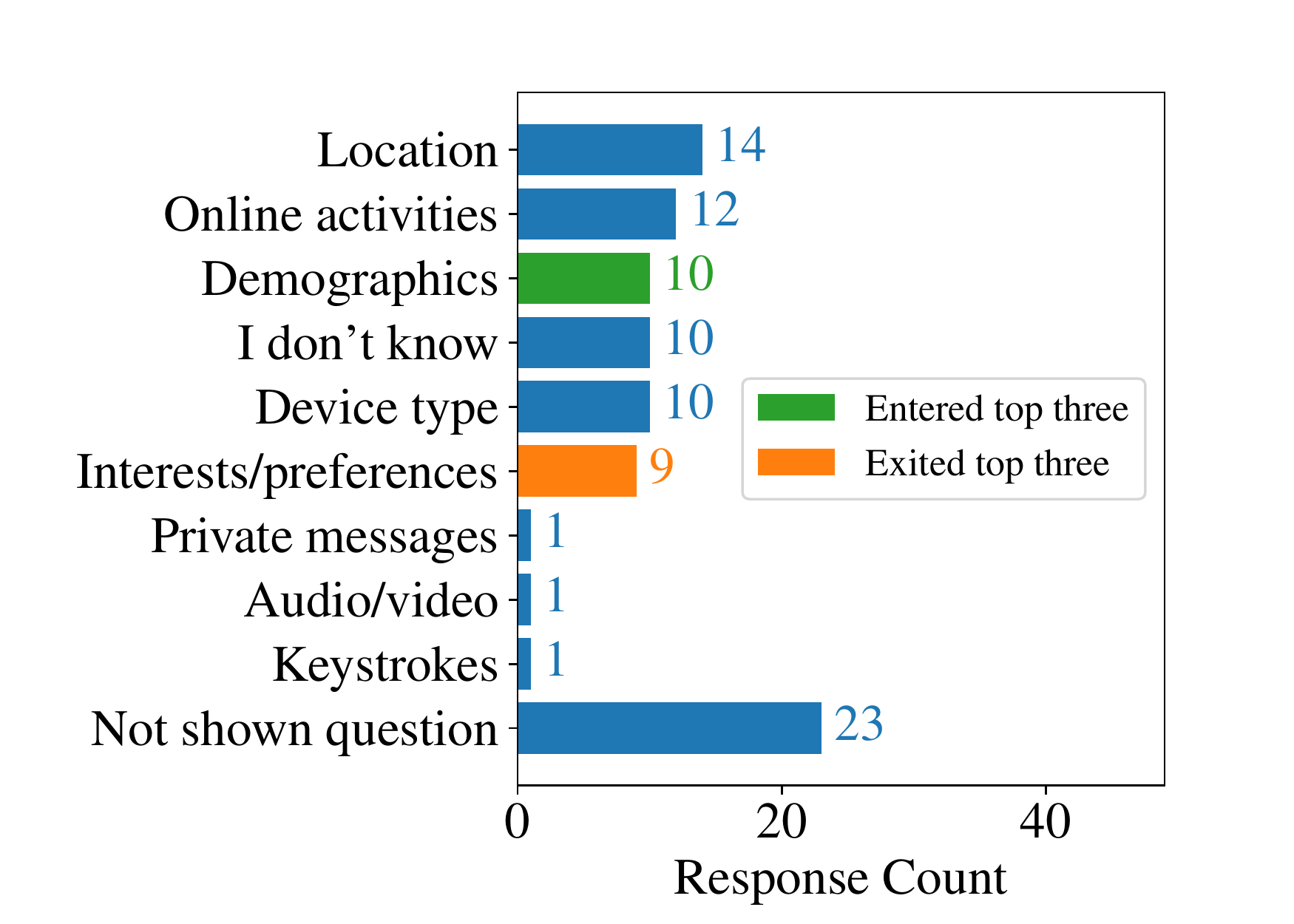}
\caption{Participants that only used university VPNs.}
\label{fig:data_shared_university} \end{subfigure}
\label{fig:q7_8_data_shared} \caption{What information do you think is being
shared with these entities? (responses selected by participants)} \end{figure}

Figure~\ref{fig:data_shared_all} shows that the 212/349 respondents that
believed other entities had access to the data collected by their VPN thought that their location (157/212), online
activities (149/212), interests (122/212), and demographic information
(121/212) were shared.  Fewer believed that private messages, recordings, or
keystrokes were shared, which coincides with what survey participants believed
VPNs could collect.

Figure~\ref{fig:data_shared_university} also shows that 14/49 survey
participants who only used VPNs provided by their university believed that
their location was being shared with third parties, and 12/49 believed their
online activities were shared.  Similarly, 37/91 of survey participants that
only used commercial VPNs believed that their location was shared, and 37/91
believed their online activities were shared.  For this survey question, there
was a significant correlation between the responses given by all participants
and participants that only used commercial VPNs, with a Spearman correlation
coefficient of 0.975 (p $\ll$ 0.001).

\subsubsection{Students do not expect anonymity from VPNs}
\paragraph{Interview} More interview participants did not believe that VPNs
guaranteed them anonymity (20/32) than those who felt the VPN did offer
privacy and anonymity (8/32). In fact, three quarters of interview
participants (24/32) told us that you can be tracked while using VPNs, and
some believed that there is always a way to do so (8/32) and that you can be
tracked by VPN provider itself~(9/32). P1 explained: \begin{quote}If it is SSL
    encryption, the VPN provider would still know that you are communicating
    with a certain web service but the VPN provider would not or probably not
    know the contents of the communication if it's SSL encrypted. They would
    only know who you want to communicate with. And if it's not encrypted,
    then they can see. They can be doing packet sniffing or even more
    malicious things like deep packet injection and deep packet inspection to
    actually look at the contents of that communication and do potential
    malicious things with that. \end{quote}

Moreover, 4/24 interview participants were convinced that the government could
track you even while using a VPN. For example, P30 used a VPN only in
different countries to access blocked content.  They did not continue to use a
VPN in the US as they did not need a VPN to access content anymore and did not
see any privacy advantages because all VPNs are \textit{``partially,
controlled or transparent to the government"}. Other interview participants
(2/24) believed that one can still be tracked by advertising agencies even if
the VPN makes tracking at least harder than normal. To overcome tracking, P21
explained that using a VPN is not enough, and they changed locations
frequently when they connected to a VPN: \begin{quote}Yes, especially if I'm using the same IP address. That
    creates a problem because my Internet footprint \dots Chrome, for
    example, my web browser could definitely still track me and connect that,
    see where I've been connecting from. Or Gmail could see that. Gmail always
    tells you, ``Oh, you've connected from this weird device, or from this
    location that we don't recognize.'' So I think you can definitely still be
tracked [while using a VPN].\end{quote}

\begin{figure}[t] \centering \begin{subfigure}[t]{0.49\textwidth}
\includegraphics[width=0.85\columnwidth]{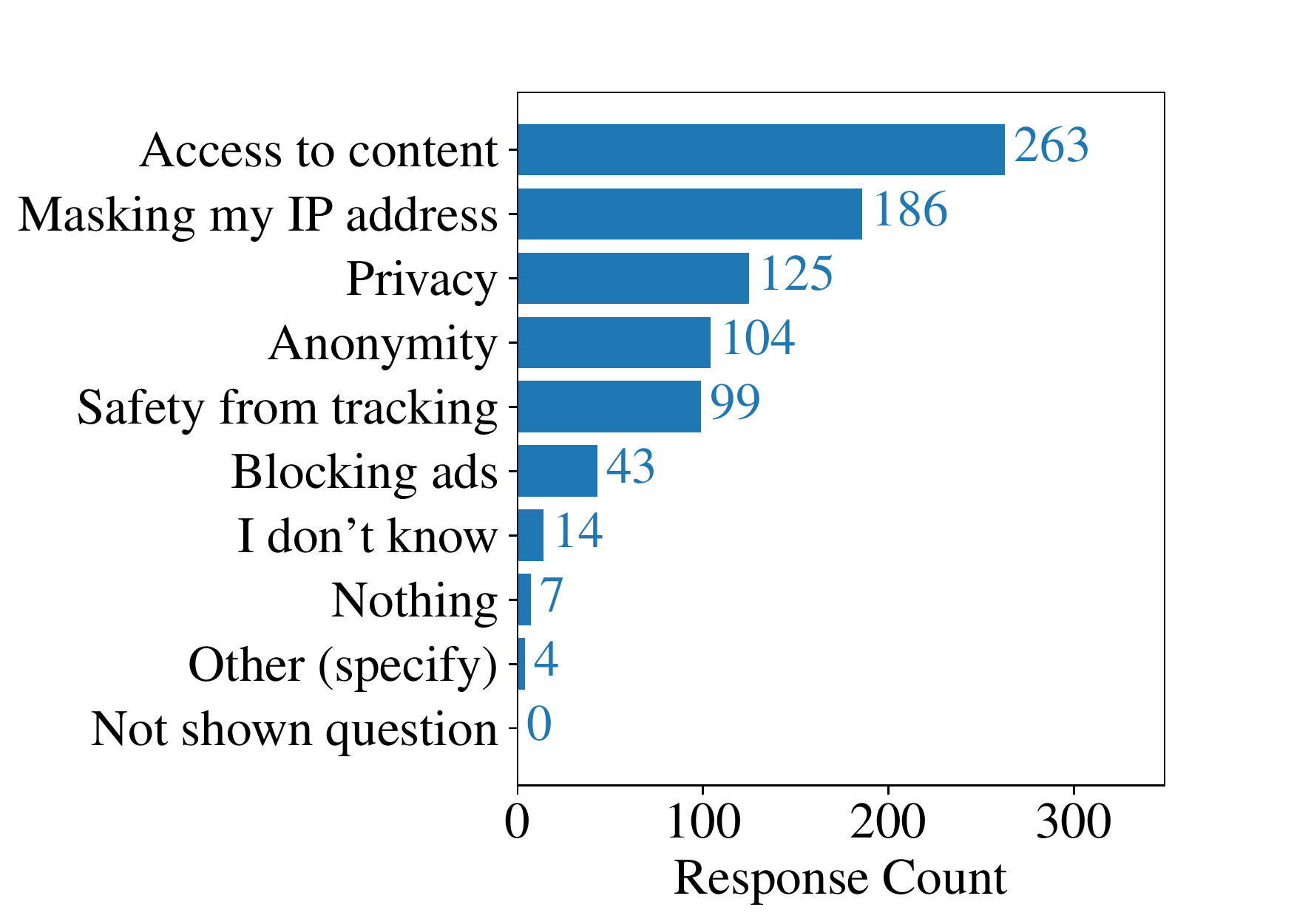}
\caption{All participants.} \label{fig:vpn_guarantees_all} \end{subfigure}
\hfill \begin{subfigure}[t]{0.49\textwidth}
\includegraphics[width=0.85\columnwidth]{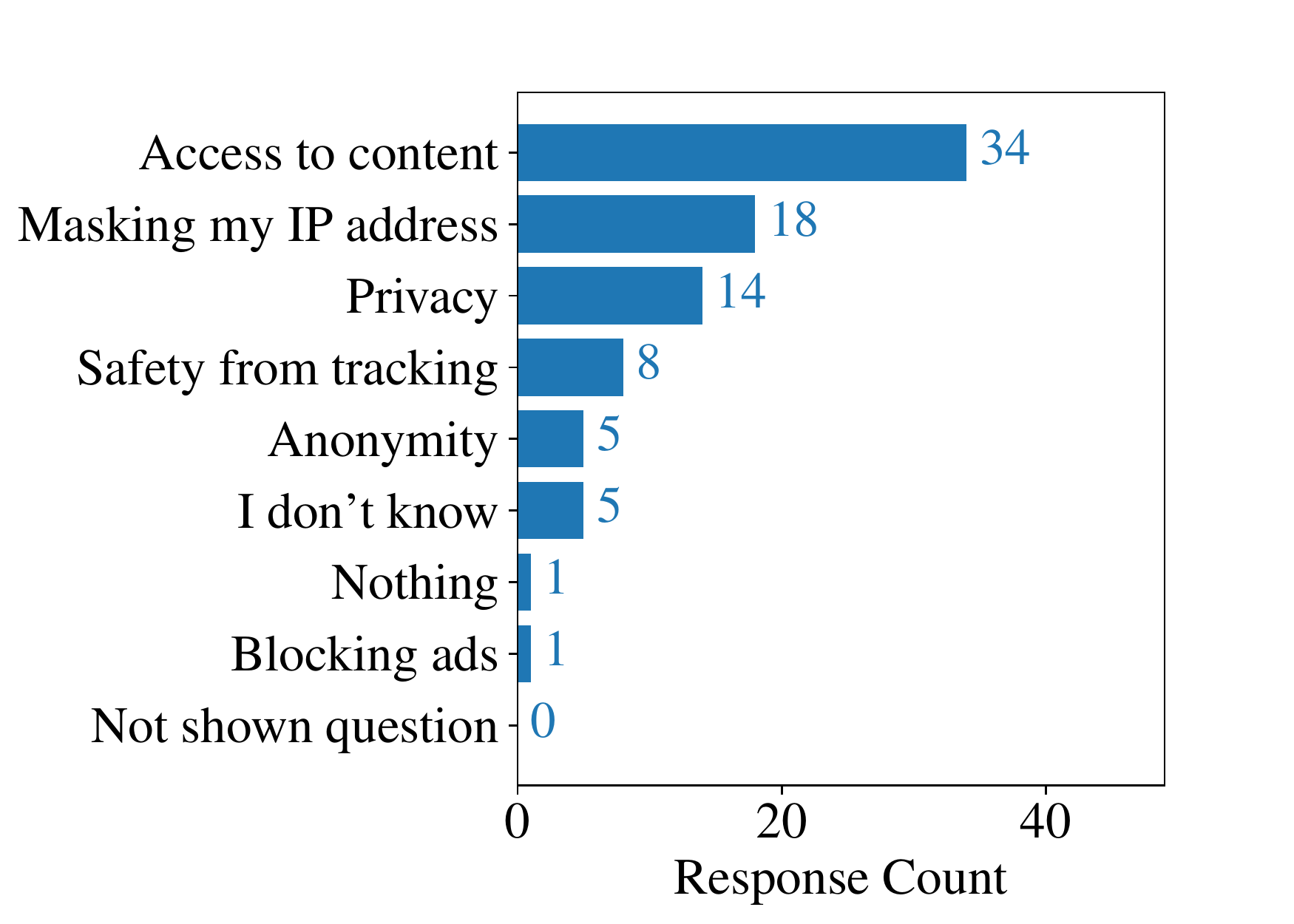}
\caption{Participants that only used university VPNs.}
\label{fig:vpn_guarantees_university} \end{subfigure}
\label{fig:q7_10_vpn_guarantees} \caption{What do you think your VPN
guarantees? (responses selected by participants)} \end{figure}

\paragraph{Survey} Figure~\ref{fig:vpn_guarantees_all} shows that survey
participants generally did not feel that VPNs provided total anonymity. Three
quarters of all participants believed their VPNs guaranteed access to certain
content (263/349)  and masking of their IP addresses (186/349). However, only
around one third of survey respondents believed that their VPNs guaranteed
privacy (125/349), anonymity (104/349), and safety from tracking (99/349).

As Figure~\ref{fig:vpn_guarantees_university} shows, we see similar results
for students that only use VPNs provided by their university.  These
participants most commonly believed that VPNs guarantee access to content
(34/49), IP address masking (18/49), and privacy (14/49).  A smaller
proportion of these students believed that their VPN guarantee them anonymity
(5/49).  Students that only used commercial VPNs most commonly believed that
VPNs guarantee access to content (65/91), IP address masking (55/91), and
anonymity (43/91). A smaller proportion of these students believed that VPNs
guarantee privacy (40/91).  For this survey question, there was a significant
correlation between the responses given by all participants and participants
that only used commercial VPNs, with a Spearman correlation coefficient of
0.954 (p $\ll$ 0.001).

    \section{Discussion}\label{sec:discussion}

Students may be surrendering {\em more} privacy through
their choice of a VPN than had they never opted to use one on the first
place~\cite{khan_1}.  This state of affairs suggests several directions for
future work concerning efforts to both raise awareness: (1) raising
awareness about data collection; (2) improving inference literacy (i.e.,
understanding about what can be inferred from the data that is collected); (3)
and generally improving students' mental models of VPN function and operation. 

\paragraph{Raising awareness about VPN data collection.}
Many students are unconcerned about
whether VPNs collected data about them but
the same students often said that 
they used a VPN to protect their data from ``companies''.  This logical
discontinuity suggests that many students may have fundamental
misunderstandings about how data is collected about them on the Internet, and
which ``companies'' are capable of doing so. 
For example, Hotspot Shield---a commercial VPN with millions of
users---leaks information about users to several trackers.  Any time a user
visits a website that includes trackers hosted on
\texttt{www.google-analytics.com}, \texttt{pixel.quantserve.com}, or
\texttt{event.shelljacket.us}~\cite{windscribe_hotspot}.  

\paragraph{Improving ``inference literacy''.} Improved inference literacy
(i.e., general understanding about how data about a user can be used to infer
higher-order behavior, from web browsing to patterns of life) could help
students make more informed decisions about when and how they use VPNs, as
well as which VPNs to use.  Students may benefit from concrete demonstrations
and examples of what can be inferred from data that VPNs intentionally leak to
third parties.  A tool (e.g., a browser extension) could help students
understand (1)~what data VPNs may collect about them; (2)~what data leaks {\em
outside} of the VPN (e.g., to ISPs, content providers), ultimately improving
their ability to select a VPN provider.  

\paragraph{Improving mental models of VPN function.}
Many students do not understand how VPNs work,
which can make it difficult for them to evaluate the security and
privacy characteristics of a VPN.
Mental models of VPN functionality could be improved with design elements that better
communicate the actions of the VPN, which could additionally improve user
perception of VPN transparency. 
Intuitive graphics and tutorials could show
how a VPN manipulates and re-routes a user's traffic. Such
instructional material could make it clear both how VPNs operate and the
parties that still have access to private or sensitive user data.

    \balance\section{Conclusion}\label{sec:conclusion}

This paper explored how university students choose VPNs, how they use them,
and their attitudes about data collection by VPN providers.
We discovered that (1)~many students chose VPNs primarily to circumvent
network controls or to access content; (2)~concerns about privacy were
important but secondary to accessing content; (3)~students generally
understand what VPNs enable them to do but not much about how they work; and
(4)~students generally expected their VPN provider to be collecting data about
them.  Students valued privacy to some
extent, but their choice of VPN software and provider was ultimately a pragmatic
question of content access, performance, and cost.  
Students  could benefit from tools and
interventions that explain the risks that they assume when choosing and using
a VPN.
Future work could attempt
to repeat this study with other populations and design interventions to help
students understand these risks.

\vfill


\pagebreak
\small
\setlength{\parskip}{-1pt}
\setlength{\itemsep}{-1pt}
\balance\bibliography{paper}
\bibliographystyle{plain}



\end{sloppypar}
\end{document}